\documentclass{aa}
\usepackage[pdftex]{graphicx}
\usepackage{epsfig}
\usepackage{amsmath}
\usepackage{wasysym}
\usepackage{txfonts}
\pdfoutput=1
\usepackage[pdftex]{color,xcolor}
\usepackage[pdftex]{graphicx}
\usepackage[backref]{hyperref}               
\usepackage{hyperref}
\hypersetup{pdfauthor=Paul Molliere}
\hypersetup{backref=true, pagebackref=true, hyperindex=true, breaklinks=true,colorlinks=true,urlcolor=blue, linkcolor=blue,citecolor=blue,pagecolor=red, bookmarks=true, bookmarksopen=true}
\hypersetup{backref=true}
\usepackage{epstopdf}

\usepackage{lipsum}
\usepackage{natbib}
\usepackage{mathtools}

\bibpunct{(}{)}{;}{a}{}{,} 
\usepackage{mhchem}

\def\mearth{{\rm M}_\oplus}

\def\f1{f_{\rm I}}
\def\mj{{\rm M}_{\textrm{\tiny \jupiter }}}

\newcommand{\rj}{{\rm R}_{\textrm{\tiny \jupiter}}}

\def\beq{\begin{equation}}
\def\eeq{\end{equation}}

\def\t2{\tau_{\rm II}}

\def\sigmas0{\Sigma_{\rm s,0}}

\def\petit{\emph{petitCODE}\ }

\def\({\left(}
\def\){\right)}
\def\<{\left<}
\def\>{\right>}

\newcommand{\rch}[1]{#1}

\begin{document}

\title{Observing transiting planets with \emph{JWST}}
\subtitle{Prime targets and their synthetic spectral observations}

\author{P. Molli\`{e}re\inst{1}  \and R. van Boekel\inst{1} \and J. Bouwman
\inst{1}  \and Th. Henning\inst{1} \and P.-O. Lagage\inst{2,3} \and M. Min
\inst{4,5}}

\institute{Max-Planck-Institut f\"ur Astronomie, K\"onigstuhl 17, D-69117 
Heidelberg, Germany \and
Irfu, CEA, Universit\'e Paris-Saclay, F-9119 Gif-sur Yvette, France \and AIM, Universit\'e Paris Diderot, F-91191 Gif-sur-Yvette, France \and SRON 
Netherlands Institute for Space Research, Sorbonnelaan 2, 3584 CA Utrecht, 
The Netherlands \and Astronomical institute Anton Pannekoek, University of 
Amsterdam, Science Park 904, 1098 XH Amsterdam, The Netherlands}

\offprints{Paul MOLLIERE, \email{molliere@mpia.de}}

\date{Received  / Accepted }

\abstract
{The \emph{James Webb Space Telescope} will enable astronomers to obtain exoplanet spectra of unprecedented precision. Especially the \emph{MIRI} instrument may shed light on the nature of the cloud particles obscuring planetary transmission spectra in the optical and near-infrared.}
{We provide self-consistent atmospheric models and synthetic \emph{JWST} observations for prime exoplanet targets in order to identify spectral regions of interest and estimate the number of transits needed to distinguish between model setups.}
{We select targets which span a wide range in planetary temperature and surface gravity, ranging from super-Earths to giant planets and have a high expected SNR. For all targets we vary the enrichment, C/O ratio, presence of optical absorbers (TiO/VO) and cloud treatment. We calculate atmospheric structures and emission and transmission spectra for all targets and use a radiometric model to obtain simulated observations. We analyze \emph{JWST}'s ability to distinguish between various scenarios.}
{We find that in very cloudy planets such as GJ~1214b less than 10 transits with \emph{NIRSpec} may be enough to reveal molecular features. Further, the presence of small silicate grains in atmospheres of hot Jupiters may be detectable with a single \emph{JWST MIRI} transit. For a more detailed characterization of such particles less than 10 transits are necessary. Finally, we find that some of the hottest hot Jupiters are well fitted by models which neglect the redistribution of the insolation and harbor inversions, and that 1-4 eclipse measurements with \emph{NIRSpec} are needed to distinguish between the inversion models.}
{Wet thus demonstrate the capabilities of \emph{JWST} for solving some of the most intriguing puzzles in current exoplanet atmospheric research. Further, by publishing all models calculated for this study we enable the community to carry out similar or retrieval analyses for all planets included in our target list.}

\titlerunning{Prime targets for JWST exoplanet observations}
\authorrunning{P. Molli\`ere et al.}

\maketitle

\section{Introduction}
\label{sect:intro}
The \emph{James Webb Space Telescope} (\emph{JWST}), will be one of the 
most exciting instruments for exoplanet science in the years to come. It will be 
the first telescope to offer a continuous wavelength coverage from 0.6 to 28 $
\mu$m \citep{beichmanbenneke2014}. Unaffected by telluric absorption, it will 
allow to probe the red optical part of a planetary spectrum (including the K-doublet 
line in hot jupiters), the near-infrared (NIR) (including molecular transitions of water and 
methane) and the mid-infrared (MIR) part.

Observing exoplanet transit spectra in the MIR may be key 
\citep{wakefordsing2015} to identify cloud species which often weaken or even 
fully blanket the atomic and molecular features in the optical and NIR, see, e.g., 
GJ1214b \citep{kreidbergbean2014},  HD 189733b \citep{sing2011}, WASP-6b 
\citep{jordan2013}  and WASP-12b \citep{singetangs2013}.

This claim for cloudiness has often been based on lacking or muted alkali 
absorption features and strong Rayleigh signatures in the optical part of hot jupiter 
transmission spectra, or muted water features in the NIR. Additionally entirely 
featureless transmission spectra (GJ~1214b) have been 
observed.

While it was theorized that muted water features could also be caused by 
depletion of water \citep{madhusudhancrouzet2014}, or a general depletion of 
metals in the atmospheres, evidence nowadays seems to point to the 
presence of clouds \citep{singfortney2015,iyerswain2016} \rch{or a combination of clouds and metal depletion \citep{Barstow:2016wa}}.

At the same time the nature of these clouds is still not known. Depending on 
the size of the cloud particles their opacity in the optical and NIR transitions 
from a Rayleigh slope (small particles) to a flat, gray opacity (large 
particles).
The resonance features of possible chemical equilibrium cloud 
species all lie in the MIR \citep{wakefordsing2015} such that a distinction 
between cloud species may only be possible within the MIR region.
The formation of clouds and hazes by non-equilibrium processes is another 
possibility, although especially hot jupiters seem to be too hot for the 
``classical'' case of hydrocarbon hazes \citep{liangseager2004} as well as for newly suggested pathways such as photolytic sulfur clouds \citep{zahnlemarley2016}.

The predicted data quality in combination with the wavelength coverage of 
\emph{JWST} will enable exoplanet atmosphere characterization to a degree 
which is, using todays observational facilities, impossible. The possible gain when having high 
quality MIR data can be seen when considering the pre-warm \emph{Spitzer} 
spectrum for HD 189733b \citep[from $\sim$ 5-14 $\mu$m, see][]
{grillmairburrows2008,todorovdeming2014} and the increased capability to 
characterize this planet's atmosphere using retrieval when compared to 
planets with sparse photometric data \citep{line2014}.

While the question of the origins of clouds is fundamental, and not answered 
yet, future efforts of characterizing planetary atmospheres will increasingly 
concentrate on the quantitative characterization of atmospheres. Using 
emission spectra the characterization even of atmospheres appearing cloudy 
in transmission may well be possible, as it is the case for HD 189733b 
\citep{barstow2014,line2014}. The reason for this is the clouds' 
decreased optical depth when not being viewed in transit geometry 
\citep{fortney2005}. Moreover, the analysis of the resulting atmospheric composition and 
abundance ratios may allow to place constraints on the planet's formation 
location, although the exact interpretation of the atmospheric abundances 
depends on the assumptions and the degree of complexity of the model being 
used to describe the planet formation and evolution \citep{obergmurray-clay2011,ali-dibmousis2014,thiabaudmarboeuf2014,hellingwoitke2014,marboeufthiabaud2014a,marboeufthiabaud2014b,madhusudhanamin2014,mordasinivanboekel2016,obergbergin2016,madhubitsch2016,cridlandpudritz2016}.

As the launch of \emph{JWST} (currently projected for October 2018) draws 
nearer the exoplanet community is in an increased need of predictions by both 
instrument and exoplanet models in order to maximize the scientific yield of 
observations. The actual performance of \emph{JWST} will only be known 
once the telescope has been launched and the first observations have been 
analyzed \citep{stevensonlewis2016}. Nonetheless, the modeling efforts of the 
telescope performance in conjunction with models of exoplanet atmospheres 
have already been started and include 
\citet{demingseager2009,batalhakalirai2015,mordasinivanboekel2016}. 
Studies which additionally look into the question of retrievability of the 
atmospheric properties as a function of the planet-star parameters can be found 
\citet{barstoweaigrain2015,greeneline2016,barstowirwin2016}. 
\citet{barstoweaigrain2015} also included time-dependent astrophysical noise 
(starspots) for stitched observations.

In this study we present detailed self-consistent atmospheric model 
calculations for a set of exoplanets which we have identified as prime scientific targets 
for observations with \emph{JWST}. The target selection was carried out 
considering the planets' expected signal-to-noise ratio for both transit and 
emission measurements, putting emphasis on a good SNR for observations 
with \emph{JWST}'s \emph{MIRI} instrument to allow measurements in the MIR. 
The planets uniformly cover the ${\rm log}(g)$--$T_{\rm equ}$ parameter 
space which also may allow to predict the objects' cloudiness 
\citep{stevenson2016}. 
We calculate a suite of models for every candidate planet. We vary the 
planetary abundances by adopting different values for [Fe/H] and C/O, including 
very high enrichments (and high mean molecular weights) for super-Earth and 
neptune-sized planets. For all planets we additionally calculate models 
including clouds, setting the free parameters of the cloud model to produce 
either small or large cloud particles which we assume to be hollow spheres to mimic irregularly shaped dust aggregates. Alternatively we assume a spherically-homogeneous shape.
For the hottest target planets TiO and VO opacities are optionally considered.
The  irradiation is treated as either assuming a dayside or global average of the received insolation flux.
For some very hot planets we additionally calculate emission spectra neglecting any energy redistribution by winds.

For all target planets we present synthetic emission and
transmission spectral observations for the full \emph{JWST} wavelength
range and compare them any existing observational data. 

For conciseness we present an exemplary analysis for a subset of our targets considered this paper. To this end we select 3 specimen belonging to the classes of extemely cloudy super-Earths, intermediately irradiated gas giants, and extremely hot, strongly irradiated gas giants, respectively. We discuss how \emph{JWST} may shed light on the nature of these planetary classes, by detecting molecular features in the NIR transmission spectra for cloudy super-Earths, or by identifying cloud resonance features in the MIR for hot jupiters. For the hottest planets we study how well \emph{JWST} can distinguish between various models which individually fit well to the data currently available.

For all targets we publish the atmospheric structures, spectra and simulated observations online. These results will facilitate predicting the
expected signal quality of existing prime exoplanet targets for \emph{JWST}.

In Section \ref{sect:models_desc} we describe the models used for carrying out our calculations, while in Section \ref{sect:candidates_select_main} the candidate selection criteria and target list are described. In Section \ref{sect:params_select} we describe which parameter setups were considered for all planets. This is followed by a characterization of the results in Section \ref{sect:calc_res}. Next, we analyze the synthetic observations of a selected subsample of targets in Section \ref{sect:sim_obs}. Finally, we describe the extent and format of the data being published in Section \ref{sect:data_pub} and describe our summary and conclusions in Section \ref{sect:summ_and_conc}.

\section{Modelling}
\label{sect:models_desc}
\subsection{Atmospheric model}
The \emph{petitCODE} calculates the atmosphere's radiative-convective 
equilibrium structure in chemical equilibrium. The chemistry module includes 
condensation, delivers abundances for the gas opacities and serves as an 
input for the cloud model. During the computation of the atmospheric 
equilibrium structures scattering is included for solving the radiative transport 
equation. For a fully converged atmosphere the code returns the atmospheric 
structure (such as temperature, molecular abundances, etc.) as well as the 
atmosphere's emission and transmission spectrum at a resolution of $
\lambda/\Delta \lambda = 1000$, where $\Delta \lambda$ is the size of a spectral bin.

Since the first version of the code described in \citet{mollierevanboekel2015} 
we have introduced several additions which lead to the current set of capabilities 
described above. These additions have been partially described in 
\citet{mancinikemmer2016a} and \citet{mancini2016b}, but in this work we provide a full description (see Appendix \ref{app:changes}).

\begin{figure*}
\centering
\includegraphics[width=0.485\textwidth]{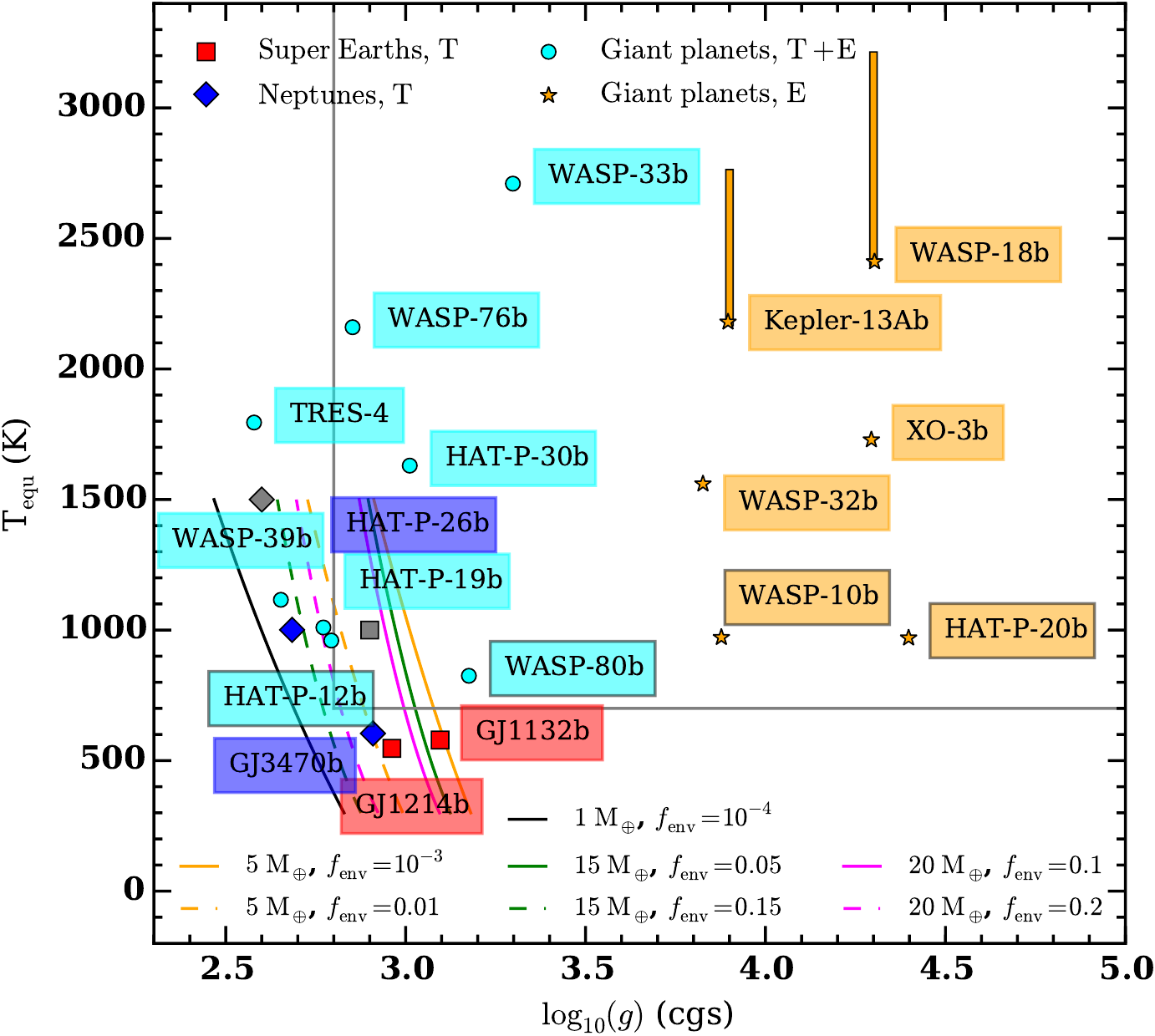}
\includegraphics[width=0.465\textwidth]{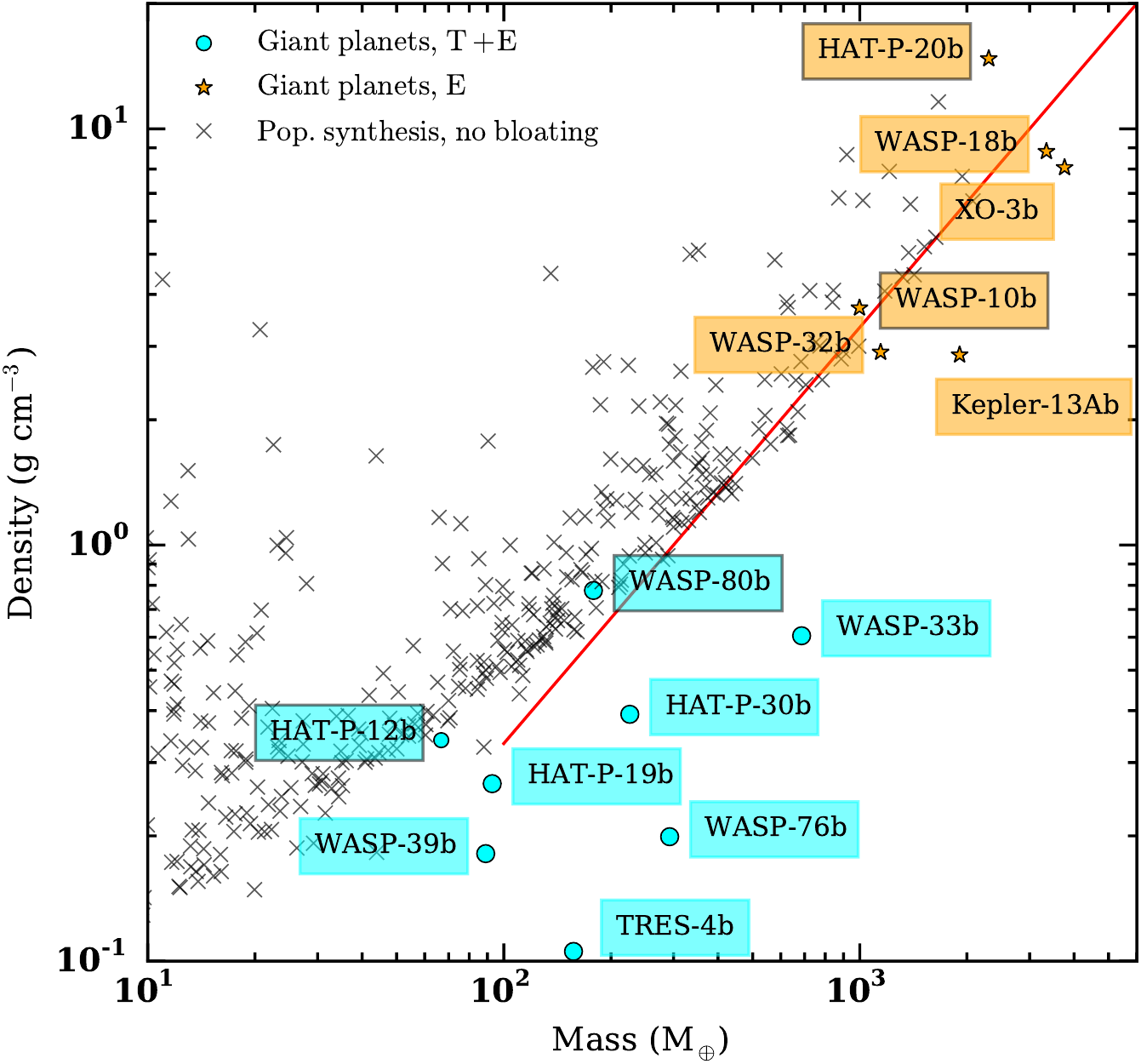}
\caption{{\it Left panel:} Our transit candidates in ${\rm log}(g)$--$T_{\rm equ}$ 
space. See the legend for the meaning of the symbols. Candidates with gray 
symbols are artificial and have been introduced to fill in the parameter space. 
``T'' and ``E'' in the legend stand for planets which can be observed in 
transmission or emission, respectively. The vertical and horizontal lines 
separate the mostly cloud-free (upper right region) from the potentially cloudy 
atmospheres (left and bottom regions) as defined in \citet{stevenson2016}. 
The black, orange, green, and magenta lines show ${\rm log}(g)$--$T_{\rm 
equ}$ curves of super-Earths and Neptune-like planets 
\citep{lopezfortney2014} for the masses and envelope mass fractions as 
described in the legend. The giant planets which have a gray frame around 
their name box are not inflated. The upper ends of the vertical orange lines shown for Kepler-13Ab and WASP-18b denote the maximum brightness temperatures observed for these planets.  {\it Right panel:} Planetary mean density as a 
function of mass. Only the giant planet candidates are plotted here, in the 
same style as in the left panel. A sample of synthetic, non-inflated planets 
calculated using the model of 
\citet{mordasinialibert2012,alibertcarron2013,jinmordasini2014} is shown as 
gray crosses. The value of the straight red line shown in the model is a linear 
function of the planetary mass, as it is expected for non-inflated giant planets 
\citep[see, e.g.][]{baruteaubai2016}.}
\label{fig:cands}
\end{figure*}

Here we summarize the additions to the code:
\begin{itemize}
\item {\it Line cutoff}: We now apply a sub-Lorentzian line treatment to all 
molecular and atomic lines far away from the line center.
\item {\it Chemistry}: We now use a self-written Gibbs minimizer for the 
equilibrium chemistry which reliably converges between 60--20,000 K and 
treats the condensation of 15 different species.
\item {\it Clouds}: We implemented the \citet{ackermanmarley2001} cloud 
model, for which we introduce a new derivation in Appendix 
\ref{app:cloud_deriv}, {\it showing that the cloud model is independent of the 
microphysics of nucleation, condensation, coagulation, coalescence and 
shattering} as long as some underlying assumptions are fulfilled. The 
implementation of the cloud model is described in Appendix \ref{app:changes}. 
We treat mixing arising from convection, convective overshoot and stellar 
irradiation.
\item {\it Cloud opacities}: For the clouds we calculate particle opacities using the
distribution of hollow spheres (DHS) or Mie theory. For this we use the code 
by \citet{minhovenier2005}. While Mie theory follows the classic assumption of spherical, homogeneous grains, DHS theory assumes a distribution of hollow spheres in order to approximate the optical properties of irregularly shaped dust aggregates. We consider clouds composed of MgAl$_2$O$_4$, Mg$_2$SiO$_4$, Fe, KCl and Na$_2$S. In our model clouds of different species cannot interact and are treated separately.
\item {\it Transmission spectra} can now be calculated.
\item The {\it molecular opacity database} has been updated to include 
Rayleigh scattering for H$_2$, He, CO$_2$, CO, CH$_4$ and H$_2$O. 
Additionally we can optionally include TiO and VO opacities.
\item{\it Scattering}: We implemented scattering for both the stellar and 
planetary light applying local accelerated lambda iteration (ALI) 
\citep{olsonauer1986} and Ng-acceleration \citep{ng1974}.
\end{itemize}

\subsection{Radiometric model}
We simulate \emph{JWST} observations with the \emph{EclipseSim} package \citep{vanboekel2012}. For this we consider observations in the \emph{NIRISS} \citep{doyonhutchings2012}, \emph{NIRSpec} \citep{	ferruitbagnasco2012}, and \emph{MIRI} instruments \citep{wrightrieke2010}. The length of each observation is taken to be the full eclipse duration, bracketed by ``baseline'' observations before and after the eclipse which have a duration of the eclipse itself each.

For the synthetic observations we assumed the instrumental resolution. If needed these spectra can be re-binned to lower resolution. For the \emph{NIRISS} slitless spectroscopy the \emph{SOSS} mode in first order will be used. For \emph{NIRSpec} we assume the \emph{G395M} mode and for \emph{MIRI} the \emph{LRS} mode. Using these modes one obtains a close to complete spectral coverage between 0.8 and 13.5 micron. However, since \emph{JWST} can only observe in one instrument at a time, one needs 3 separate observations to obtain the complete spectral coverage.

\section{Candidate selection}
\label{sect:candidates_select_main}
\subsection{Candidate selection criteria}
\label{sect:candidates_select}
In order to obtain a list of well observable candidates we only considered 
planets for which the transit times are known accurately. As a next criterion we 
checked if  a candidate is observable in transmission and/or emission with 
signal amplitude (at 7 $\mu$m) of SNR $>$ 5, where the noise is assumed to 
be photon noise + a 50 ppm noise floor. For this initial check we approximated 
the planetary emission using a blackbody spectrum, whereas for the 
transmission signal we assumed a transit signal amplitude of 5 pressure scale 
heights.

This initial analysis results in a large number of possible targets, given the wealth of transiting exoplanets known already today. In order to maximize the scientific yield of the first \emph{JWST} observations it may be instructive to first observe a planetary sample as diverse as possible and to embark on more detailed studies within a given planetary class later on. Our goal therefore is to define such a diverse target list, and to map out the parameter space defining planetary classes as well as possible.
  
The main physical parameter space for candidate selection in the work 
presented here is the ${\rm log}(g)$--$T_{\rm equ}$ space. This space is quite 
fundamental in the sense that the equilibrium temperature and the planetary 
surface gravity are two key parameters impacting the pressure temperature 
structure and spectral appearance of a planet \citep[see, e.g.][]
{sudarsky2003,mollierevanboekel2015}. The total atmospheric enrichment can 
be of importance too, but for scaled solar compositions variations of  ${\rm 
log}(g)$ and [Fe/H] are degenerate to some degree 
\citep{mollierevanboekel2015}. Additionally, for every target planet we will present calculations assuming a larger or smaller enrichment than used in the fiducial case. Further, the planet's location in the ${\rm log}(g)
$--$T_{\rm equ}$ space may allow to assess if a planet is cloudy or not 
\citep[see][]{stevenson2016}. We aimed for a broad coverage of 
candidates in our parameter space.

In addition, we divided the candidates in super-Earths, hot neptunes, inflated 
and non-inflated giant planets and tried to select a sample as diverse as 
possible, still above observational thresholds described above, however.

Our final selection is shown in the left panel of Figure \ref{fig:cands}. For the 
super-Earths and hot Neptunes we only have 4 candidates in total, 2 for each 
class. In order to estimate the region usually occupied by these two classes of 
planets we overplot ${\rm log}(g)$--$T_{\rm equ}$ lines for super-Earths and 
hot neptunes of varying mass and envelope fraction \citep[taken from][]
{lopezfortney2014} to investigate the area which super-Earths and Neptunes may 
occupy in ${\rm log}(g)$--$T_{\rm equ}$ space. We added two more hot, 
artificial candidates, one for the neptunes and one for the super-Earths to 
increase the coverage in ${\rm log}(g)$--$T_{\rm equ}$ space. We cannot 
introduce even hotter artificial super-Earths or Neptunes, because their 
envelopes would likely be evaporated \citep{jinmordasini2014}.

The giant planets were divided in inflated giant planets, non-inflated giant 
planets, and inflated giant planets associated to non-inflated giant planets. 
This association means that the inflated planets are lying closely to the non-inflated ones in ${\rm log}(g)$--$T_{\rm equ}$ space, but are inflated. It is 
known that whether or not inflation occurs is correlated with irradiation 
strength and thus $T_{\rm equ}$ 
\citep{laughlincrismani2011,demoryseager2011}. Therefore such close 
neighbors in ${\rm log}(g)$--$T_{\rm equ}$ space, showing inflation or no 
inflation, may potentially shed light on the mechanisms driving inflation.
To assess whether a giant planet is inflated we show the density of all 
candidates as a function of their mass in the right panel of Figure 
\ref{fig:cands}. We also plot a synthetic planetary population, calculated using 
the planet formation and evolution code of 
\citet{mordasinialibert2012,alibertcarron2013,jinmordasini2014}, 
which does not include any inflation processes. All planets that have a 
density lower than the one shown for the synthetic giant planets were 
considered to be inflated.

\subsection{List of selected candidates}
\label{sect:target_list}
The parameters of the exoplanet targets modeled in this paper are given in Table 
\ref{tab:obs_cands}.

For the super-Earths we included a planet with a mass of 5 $\mearth$, placed 
at a distance of 0.1~AU around a sun-like star, which corresponds to a 
planetary equilibrium temperature of 880~K. We assumed an initial H--He 
envelope mass fraction of 1\% of the planet's total mass. 
Calculations by \citet{jinmordasini2014} indicate that such a planet loses 
about half of its envelope in the first 20 Myr of its lifetime, therefore we 
calculate the radius of the planet using the relation by \citet{lopezfortney2014} 
for a 5~$\mearth$ planet with a 0.5 \% envelope fraction at an age of 20~Myr.
At later times the envelope of such a planet will be evaporated even further, therefore the high enrichment we assume for its atmosphere (see Section \ref{sect:params_select}) may be seen as a proxy for a secondary atmosphere with a high mean molecular weight. 
Note that photo-evaporation is not included in the calculations of 
\citet{lopezfortney2014}. We refrained from considering even hotter super 
Earth planets, because more strongly irradiated planets will be more strongly 
affected by photo-evaporation such that the primordial planetary H--He 
envelope may not survive.

For the Neptunes we considered an artificial object with a mass of 20 $
\mearth$ in orbit around a sun-like star with a semi-major axis of 0.05~AU, 
corresponding to an effective temperature of 1250~K. The H--He envelope 
mass fraction was taken to be 15\%. Again we consulted 
\citet{jinmordasini2014} and found that such a planet may retain a significant 
amount of its envelope up to 100 Myr and longer.

The parameters for the artificial planets listed in Table \ref{tab:obs_cands} 
were again obtained using \citet{lopezfortney2014}.

\begin{table*}[t]
\centering
\begin{tabular}{lccccccccc}
Planet name & $M_{\rm Pl}$ ($\mj$) & $M_{\rm Pl}$ ($\mearth$) & $R_{\rm 
Pl}$ ($\rj$) & ${\rm log_{10}}(g_{\rm Pl})$ (cgs) & $T_{\rm equ}^{\rm f}$  (K) & 
[Fe/H]$_*$ & [Fe/H]$_{\rm Pl}^{\rm fid}$ & Similar planets & Obs. References \\ \hline \hline
{\it Super-Earths} & & & & & & & & \\ \hline
GJ~1132b & 0.01 & 1.62 & 0.1 & 3.1 & 579 & -0.12 & 3$^{\rm a}$ & -- & -- \\
GJ~1214b & 0.02 & 6.36 & 0.24 & 2.96 & 547 & 0.39 & 3$^{\rm a}$ & -- & (1.1-7) \\
{Artificial}$^{\rm b}$ & 0.02 & 5 & 0.29$^{\rm c}$ & 2.67 & 880 & 0.00 & 3$^{\rm a}$ & -- & -- \\ \hline \hline
{\it Neptunes} & & & & & & & & \\ \hline
GJ3470b & 0.04 & 13.9 & 0.37 & 2.91 & 604 & 0.17 & 1.98 & HATS-6b$^{\rm d}$ & (2.1-6) \\
HAT-P-26b & 0.06 & 18.59 & 0.56 & 2.68 & 1001 & 0.01 & 1.56 & -- & (3.1) \\
{Artificial}$^{\rm b}$ & 0.06 & 20 & 0.65$^{\rm c}$ & 2.45 & 1250 & 0.00 & 1.51 & -- & -- \\ \hline \hline
{\it Gas giants} & & & & & & & & \\ \hline
WASP-80b & 0.56 & 178.62 & 0.99 & 3.18 & 825 & -0.14 & 0.60 & HAT-P-17b & (4.1-3) \\
HAT-P-12b & 0.21 & 66.74 & 0.94 & 2.79 & 960 & -0.29 & 0.75 & WASP-67b, & (5.1-5.4) \\
 & & & &  &  &  &  & HAT-P-18b & \\
 WASP-10b & 3.14 & 997.98 & 1.04 & 3.88 & 972 & 0.05 & 0.27 & WASP-8b$^{\rm e}$ & (6.1) \\
HAT-P-20b & 7.25 & 2302.98 & 0.87 & 4.4 & 970 & 0.35 & 0.32 & -- & (7.1)  \\ \hline \hline
{\it Inflated giants} & & & & & & & & \\ \hline
TrES-4b & 0.49 & 157.01 & 1.84 & 2.58 & 1795 & 0.28 & 1.1 & WASP-17b, & (8.1-5) \\
 & & & &  &  &  &  & WASP-94b, & \\
  & & & &  &  &  &  & WASP-79b & \\
WASP-33b & 2.16 & 686.51 & 1.68 & 3.3 & 2734 & 0.1 & 0.44 & WASP-12b & (8.5), (9.1-5) \\
HAT-P-30b & 0.71 & 225.98 & 1.34 & 3.01 & 1630 & 0.12 & 0.80 & WASP-7b & (10.1) \\
Kepler-13Ab & 6.0 & 1906.97 & 1.41 & 3.9 & 2180$^{\rm f}$ & 0.2 & 0.23 & -- & (11.1) \\
WASP-32b & 3.6 & 1144.18 & 1.18 & 3.83 & 1560 & -0.13 & 0.05 & CoRoT-2b, & (12.1) \\
 & & & &  &  &  &  & WASP-43b & \\
WASP-18b & 10.52 & 3343.55 & 1.16 & 4.3 & 2411$^{\rm f}$ & 0.1 & -0.04 & -- & (13.1) \\
XO-3b & 11.83 & 3759.9 & 1.25 & 4.29 & 1729 & -0.18 & -0.35 & HAT-P-2b & (14.1-2) \\
WASP-76b & 0.92 & 292.4 & 1.83 & 2.85 & 2160 & 0.23 & 0.84 & WASP-48b, & -- \\
 & & & &  &  &  &  & KELT-7b, & \\
 & & & &  &  &  &  & WASP-82b & \\
HAT-P-19b & 0.29 & 92.81 & 1.13 & 2.77 & 1010 & 0.23 & 1.22 & WASP-69b & (6.1), (15.1) \\
WASP-39b & 0.28 & 88.99 & 1.27 & 2.65 & 1116 & -0.12 & 0.84 & -- & (5.4), (6.1), \\
 & & & &  &  &  &  &  & (16.1-2) \\ \hline \hline
\end{tabular}
\caption{High priority targets for which we simulate spectra in this study. Note 
that for the planets listed in the ``similar planet'' section the planetary masses 
usually agree quite well to our target masses, therefore the enrichment (as 
estimated by Equation \ref{equ:enrich_pl}) may be similar. Note that the 
planetary enrichment is also linearly dependent on the host star's metallicity, 
however. Footnotes: (a): For these planets the metal mass fraction as 
estimated by Equation \ref{equ:enrich_pl} was larger than 1, such that we 
imposed a maximum metallicity value of 3. Additionally the calculations with enrichments 10 times larger than the fiducial case, which would lead to a metallicity value of 4, have been neglected for these planets. (b): These planets are artificial 
candidates in order to fill in the $T_{\rm equ}$--${\rm log}(g)$ parameter 
space. (c): The artificial super-Earth and hot Neptune have relatively large 
radii because we assumed their ages to be 20 and 100 Myr for the super 
Earth and Neptune-like planet, respectively. At later ages they would be too 
strongly affected by envelope evaporation. (d): HATS-6b is much more 
massive than GJ~3470b. Therefore only the metal depleted case (``FEH\_m
\_1'') of GJ3470b in our calculations is comparable to what we would estimate 
for HATS-6b. (e): WASP-8b is highly eccentric ($e=0.31$). (f): The equilibrium temperatures given in this table 
correspond to the values derived from the stellar and orbital parameteres, i.e. 
$T_{\rm equ} = T_*\sqrt{R_*/2a}$, where $T_*$ is the stellar effective 
temperature, $R_*$ the stellar radius and $a$ the planet's semi-major axis. 
Note that Kepler-13Ab and WASP-18b have emission brightness 
temperatures hotter than even the dayside averaged effective temperatures. 
For these planets calculations at even higher temperatures were carried out, 
see Section \ref{sect:irrad_treatment} for more information. References for 
observational data: (1.1): \citet{beanmiller2010}, (1.2): \citet{desertbean2011}, 
(1.3): \citet{beandesert2011}, (1.4): \citet{bertacharbonneau2012}, (1.5): 
\citet{frainedeming2013}, (1.6): \citet{kreidbergbean2014}, (1.7): 
\citet{caceresbabath2014}, (2.1): \citet{crossfieldbarman2013}, (2.2): 
\citet{demorytorres2013}, (2.3): \citet{nasciembipiotto2013}, (2.4): 
\citet{biddlepearson2014}, (2.5): \citet{ehrenreichbonfils2014}, (2.6): 
\citet{dragomirbenneke2015}, (3.1): \citet{stevensonbean2016}, (4.1) 
\citet{fukuikawashima2014}, (4.2)  \citet{mancinisouthworth2014}, (4.3) 
\citet{triaudgillon2015}, (5.1): \citet{lineknutson2013}, (5.2): 
\citet{todorovdeming2013}, (5.3): \citet{mallonnasciembi2015}, (5.4): 
\citet{singfortney2015}, (6.1): \citet{kammerknutson2015}, (7.1): 
\citet{demingknutson2015}, (8.1): \citet{knutsoncharbonneau2009}, (8.2): 
\citet{chaningemyr2011}, (8.3): \citet{ranjancharbonneau2014}, (8.4): 
\citet{sozzettibonomo2015}, (8.5): \citet{turnerpearson2016}, (9.1): 
\citet{smithanderson2011}, (9.2): \citet{demingfraine2012}, (9.3): 
\citet{demooijbrogi2013}, (9.4): \citet{haynesmandell2015}, (9.5): 
\citet{vonessenmallonn2015}, (10.1) \citet{fosterharrington2016}, (11.1): 
\citet{shporerorourke2014}, (12.1) \citet{garlandharrington2016}, (13.1): 
\citet{nymeyerharrington2011}, (14.1): \citet{wongknutson2014}, (14.2): 
\citet{machalekgreene2010}, (15.1): \citet{mallonvonessen2015}, (16.1): 
\citet{fischerknutson2016}, (16.2): \citet{ricciramonfox2015}}
\label{tab:obs_cands}
\end{table*}

\section{Selection of planet parameters}
\label{sect:params_select}
For every target identified in Section \ref{sect:target_list}, and modeled in 
Section \ref{sect:calc_res}, we calculate a fiducial model, and then vary five 
parameters within a parameter space which we will describe in the following.
The parameters which are studied are the atmospheric enrichment (see Section \ref{sect:params_select:atmo_enrich}), clouds (Section \ref{sect:params_select:clouds}), the C/O number ratio (Section \ref{sect:params_select:co}), the inclusion of optical absorbers in the form of TiO/VO (Section \ref{sect:params_select:tiovo}), and the redistribution of the stellar irradiation energy (Section \ref{sect:irrad_treatment}).

\subsection{Atmospheric enrichment}
\label{sect:params_select:atmo_enrich}
The analysis of the radii of known, cool ($T_{\rm equ}<1000 \ {\rm K}$) 
exoplanets using planetary structure models suggests that the planets' 
enrichment in heavy elements, $Z_{\rm Pl}$, is proportional to the host stars' 
metal enrichment $Z_{*}$. The ratio $Z_{\rm Pl}/Z_{*}$ is a function of the 
planetary mass and decreases with increasing planetary mass 
\citep{millerfortney2011,thorngrenfortney2016}.
A fit of the function
\beq
\frac{Z_{\rm Pl}}{Z_{*}} = \beta \left(\frac{M_{\rm Pl}}{\mj}\right)^\alpha
\label{equ:enrich_pl}
\eeq
to the sample of planets investigated in \citet{millerfortney2011} yields $
\alpha=-0.71\pm0.10$ and $\beta=6.3\pm 1.0$ \citep{mordasiniklahr2014}. In 
the same paper \citep{mordasiniklahr2014} fit the results of a synthetic 
population of planets formed via the core accretion paradigm and find $
\alpha=-0.68$ and $\beta=7.2$, which fits the observational data and the Solar 
System ice and gas giants. A comprehensive summary of observational evidence further backing the finding that lower mass planets are more heavily enriched than more massive planets can be found in \citet{mordasinivanboekel2016}.

An important question is to which extent the metal content of the planet is 
mixed into its envelope and atmosphere. It is suggested that for Saturn nearly all 
metals are locked into the central core, whereas for Jupiter the metals appear 
to be fully mixed into the envelope \citep{fortneynettelman2010}.
In our fiducial models we will assume planets where half of the metal 
enrichment is mixed into the planet's envelope and atmosphere.

For the fiducial models of the planets whose atmospheres we will simulate we 
thus use Equation \ref{equ:enrich_pl} to describe the atmospheric enrichment, 
taking into account an additional factor 1/2 to relate the atmospheric enrichment to the 
planetary bulk enrichment.
We will take the host star's [Fe/H] as a proxy for the stellar enrichment.

Additionally we will consider models with 10 times more or less metal 
enrichment than in the fiducial model.

\subsection{Clouds}
\label{sect:params_select:clouds}
For every planet we consider 9 different cloud model parameter setups in 
order to test a broad range of possible cloud properties.
These setups are listed in Table \ref{tab:cloud_models}.

\begin{table}[t]
\centering
\begin{tabular}{cccccccc}
Model & Shape & $f_{\rm sed}$ & $\sigma$ & $X_{\rm max}$ ($Z_{\rm Pl}$) & $a$ ($\mu$m) & Fe & SCC \\ \hline \hline
1 & DHS & 3 & 2 & -- & -- & yes & yes \\
2 & DHS & 1 & 2 & -- & -- & yes & yes \\ \hline
3 & DHS & 0.3 & 2 & -- & -- & yes & yes/no \\
4 & DHS & 0.01 & 2 & -- & -- & yes & yes/no \\ \hline \hline
5 & DHS & -- & -- & $10^{-2}$ & 0.08 & no & yes \\ 
6 & DHS & -- & -- & $3\times10^{-4}$ & 0.08 & no & yes \\
7 & DHS & -- & -- & $3\times10^{-5}$ & 0.08 & no & yes \\ \hline
8 & DHS & -- & -- & $3\times10^{-4}$ & 0.08 & yes & yes \\ \hline
9 & Mie & -- & -- & $3\times10^{-4}$ & 0.08 & no & yes \\ \hline \hline
\end{tabular}
\caption{Cloud models studied for all planetary candidates listed in Table 
\ref{tab:obs_cands}. The ``shape'' column describes whether the grain 
opacities are described assuming irregular grains (using DHS) or as 
homogeneous, spherical grains (using Mie theory). $f_{\rm sed}$ is the 
standard settling parameter from the \citet{ackermanmarley2001} cloud model 
and $\sigma$ is the width of the log-normal particle size distribution function in 
this model. Note that the value of $\sigma=1$ formally corresponds to a Dirac 
delta function. For the parametrized cloud model $X_{\rm max}$ describes the 
maximum cloud mass fraction within the atmosphere, while $a$ denotes the 
mono-disperse particle size. ``Fe'' denotes whether iron clouds have been 
included. The column ``SCC'' (``self-consistent coupling'') denotes whether the 
cloud opacities have been coupled to the atmospheric temperature iteration or 
whether the converged, cloud-free atmospheric temperature structure has 
been used to obtain spectra including clouds. Note that for, e.g., GJ~1214b models 3 and 4 converged with self-consistent coupling.}
\label{tab:cloud_models}
\end{table}
Models 1 and 2 use the \citet{ackermanmarley2001} cloud model to couple the 
effect of clouds self-consistently with the atmospheric temperature iteration. 
The values for the settling parameter $f_{\rm sed}$, which is the ratio of the 
mass averaged grain settling velocity and the atmospheric mixing velocity, 
have been adopted covering the lower range of what is typically being used 
for brown dwarfs \citep[$f_{\rm sed}=1$-$5$, see][]
{saumonmarley2008,morleyfortney2012} and we use $f_{\rm sed}=1,3$ here. Further, we account for the fact that 
Earth high altitude clouds are well described using small $f_{\rm sed}<1$ 
values and that the flat transmission spectrum of GJ~1214b is best described 
using $f_{\rm sed}\ll 1$ 
\citep{ackermanmarley2001,morleyfortney2013,morleyfortney2015}. We therefore use such small $f_{\rm sed}$ value cloud setups in models 3 and 4, \rch{namely $f_{\rm sed}$~=~0.3 and 0.01}. Similar to 
\citet{morleyfortney2015} we find that for the cases with small $f_{\rm sed}
<1$, with the planets often being quite strongly enriched, it \rch{can be} challenging 
to obtain converged results when self-consistently coupling the cloud model to 
the radiative-convective temperature iteration. Thus, for cloud model setups 3 
and 4 we follow \rch{a two-pronged approach: first, we attempt to calculate the atmospheric structures self-consistently. If this does not succeed we follow \citet{morleyfortney2015} and calculate cloudy spectra for these two model setups using the temperature structure of the fiducial, cloud-free model.}

For the cases where the cloud models 3 and 4 converged, and for all other cloud models considered here, the clouds are coupled
to the atmospheric structure iteration self-consistently.
For, e.g., GJ~1214b, which has an enrichment of 1000~$\times$
solar \rch{ in our fiducial setup}, the structures for cloud models 3 and 4 with self-consistent coupling converged. We will look at this planet in greater detail in Section \ref{subsubsect:gj1214b}.
For cases for which the self-consistent coupling between cloud models 3 and 4 and the temperature iteration 
converged we compared the resulting spectra to the calculations which 
applied models 3 and 4 to the cloud-free temperature structure. We found that 
the transmission spectra can agree quite well but may be offset due to different temperatures in the atmospheres. If the atmospheric temperatures are close to a chemically important temperature range, e.g. close to the temperature where carbon gets converted from methane (lower temperatures) to CO (larger temperatures), the transmission spectra can be quite different, with the cooler, not self-consistently coupled atmospheres exhibiting methane features which the self-consistent atmospheres lack. \rch{Analogously} emission spectra may share a similar spectral shape (not in all cases, due to the same reasons as outlined above for transmission spectra) but have a different flux normalization:
The self-consistent models 3 and 4 conserve the flux, while the post-processed cloud calculations,
simply applying clouds to the clear atmospheric structures for the spectra, do not.

We want to stress that our implementation of the \citet{ackermanmarley2001} 
cloud model differs from the version described in the original paper in two 
ways: (i) we account for vertical mixing induced by insolation, see Appendix 
\ref{sect:app_cloud_mod}. A similar approach was taken for GJ~1214b in the 
study by \citet{morleyfortney2015}. (ii) The mixing length in our cloud model 
implementation is equal to the atmospheric pressure scale height, while in the 
\citet{ackermanmarley2001} model the mixing length in the radiative layers is 
up to 10 times smaller than the atmospheric pressure scale height. This 
means that, for a given $f_{\rm sed}$ value, our clouds will be more extended, 
because the cloud density above the cloud deck is proportional to $P^{f_{\rm 
sed}/\lambda}$, where $\lambda$ is the ratio of the mixing length $L$ divided 
by the pressure scale height $H$. Further, the mixing velocity is equal to 
$K_{zz}/L$, where $K_{zz}$ is the atmospheric eddy diffusion coefficient, 
meaning that for a given $f_{\rm sed}$ value our grains will be smaller. Both 
effects effectively lower our $f_{\rm sed}$ value in comparison to the 
\citet{ackermanmarley2001} value.
See Appendix \ref{sect:app_cloud_mod} for a description of our 
implementation of the \citet{ackermanmarley2001} cloud model.

In our standard case the cloud particles are assumed to be irregularly shaped dust aggregates which we describe using the Distribution of Hollow Spheres method (DHS). This is in contrast to the case of 
homogeneous spheres in Mie theory. We investigate the effect of Mie 
opacities as a non-standard scenario in cloud model 9. Only the small cloud 
particle case is studied with Mie theory, as only then differences between DHS 
and Mie in the cloud resonance features may be seen: for larger particles the 
cloud opacity is gray for both the DHS and Mie treatment, without any 
observable features. 

So far the use of Mie theory is a standard approach for cloud particles in 
brown dwarf / exoplanet atmospheres \citep[see, e.g.,][]
{hellingackerman2008,madhusudhanburrows2011,morleyfortney2012,benneke2015,baudinobezard2015}, which is a useful starting point to assess the first 
order effect of clouds on planetary structures and spectra. However, in all 
cases where crystalline features of silicate grains have been observed in an 
astrophysical context so far it was found that the opacity of Mie grains poorly 
fits the observations. Only the use of non-homogeneous or non-spherical 
shapes, such as Distribution of Hollow Spheres (DHS) or Continuous 
Distribution of Ellipsoids (CDE) provides a good fit to the data. Examples are 
the features of dust particles in disks around Herbig Ae/Be stars 
\citep{bouwmanmeus2001,juhaszbouwman2010}, as well as AGB stars, post-AGB stars, planetary 
nebulae, massive stars, but also stars with poorly known evolutionary status 
\citep[for a discussion of the data and the spectral fits see][respectively]
{molsterwaters2002,minhovenier2003}. Given the observational evidence in 
different astrophysical scenarios we therefore chose the DHS treatment 
of grains as our standard scenario. We note that for brown dwarfs an explicit 
detection of a cloud feature is still missing \citep[although tentative evidence 
exists, see][]{cushingroellig2006} and that transmission spectra of planets 
have so far only probed the (often cloudy/hazy) optical and NIR regions which 
are devoid of cloud resonance features because these primarily lie in the MIR.
 
For models 5 to 9 in Table \ref{tab:cloud_models} we introduce a 
parametrized cloud model which corresponds to vertically homogeneous 
clouds, with the cloud mass fractions per species equal to the values derived 
from equilibrium chemistry, but not larger than $X_{\rm max}=10^{-2}\cdot Z_{\rm Pl}$, $3\times10^{-4}\cdot Z_{\rm Pl}$ or 
$3\times10^{-5}$ $\cdot Z_{\rm Pl}$, where $Z_{\rm Pl}$ is the atmospheric 
metal mass fraction. \rch{In that sense $X_{\rm max}$ can be thought of as a proxy for the settling strength, where smaller $X_{\rm max}$ values correspond to a stronger settling.}
The cloud particle radius for all grains in these models is 
fixed at 0.08 $\mu$m. For the standard setup of these clouds the contribution 
of iron clouds was neglected. Only condensed species which can exist in 
thermochemical equilibrium within the atmospheric layers were considered 
and only if the condensation-evaporation boundary was within the simulated 
domain. For every species only a single cloud layer is allowed, implicitly 
assuming that the lowest possible cloud layer for a given species traps the 
cloud forming material.

We introduced the parametrized cloud model because we found that it is only 
possible to reproduce the steep Rayleigh slope observed for some hot 
Jupiters from the optical to the near IR \citep[to $\sim1.3$~$\mu$m, see][]
{singfortney2015} if one places small cloud particles within the radius range ($
\sim$0.06 to 0.12~$\mu$m) in the high layers of the atmosphere. While the 
upper particle radius boundary results from the requirement to have a 
Rayleigh like scattering opacity down to optical wavelengths, the lower radius 
boundary results from the requirement to have a Rayleigh-like extinction out to 
the NIR. For Mg$_2$SiO$_4$, we found that the NIR extinction would become flatter for 
particles smaller than 0.06 $\mu$m, either because of absorption or 
scattering. Similar ($\sim$~0.1~$\mu$m) particle sizes have been found by 
\citet{pontsing2013,leeirwin2014,singwakeford2015a}, who report that they 
need small cloud particles at high altitudes with sizes between 0.02 and 0.1 $
\mu$m to reproduce the strong Rayleigh \rch{signal} observed in the optical and UV 
of the planets HD 189733b and WASP-31b. \rch{The need for submicron-sized cloud particles in hot jupiters has recently also been pointed out by \citet{Barstow:2016wa}, at least in certain equilibrium temperature ranges.} \citet{pontsing2013} and 
\citet{leeirwin2014} find that this small particle cloud layer may be 
homogeneous over multiple scale heights in HD 189733b. 
\citet{leeirwin2014} analyzed HD189733b by retrieving the cloud properties 
(size and optical depth) and molecular abundances and found that the need 
for a small ($<$~1~$\mu$m) cloud particles is a robust finding, independent 
from variations of the planet's radius, terminator temperature and cloud 
condensate species. Moreover, this small cloud particle size is consistent with 
the lower boundary of grain sizes derived in \citet{parmentierfortney2016} 
when studying the optical phase curve offsets of hot jupiters.

Iron clouds are neglected for the cloud models 5-7 and 9 because the 
strongly absorbing nature of iron in the optical does not allow for the 
\rch{dominance} of Rayleigh scattering in the optical. For illustrative reasons the 
case were iron opacities were included in the small particle regime has been 
studied in model 8.

\rch{For the cool super-Earths, Neptunes and coolest planets in general (GJ~3470b, HAT-P-26b, GJ~1214b, GJ~1132b and WASP-80b) only Na$_2$S and KCl are considered as possible cloud species, because it is doubtful that higher temperature condensates can be mixed up from the deep locations of their cloud decks \citep{charnaymeadows2015,parmentierfortney2016}. For the planets which are only slightly hotter (WASP-39b, HAT-P-19b, HAT-P-12b, WASP-10b and HAT-P-20b) we consider both cases using either the full condensate or only the Na$_2$S and KCl condensate model, where the models including only Na$_2$S and KCl may be more appropriate in this temperature regime.}

\subsubsection{Crystalline or amorphous cloud particles?}
Throughout this work we will assume crystalline cloud particles, rather than 
amorphous ones, as long as the corresponding optical data are available. 
This is a very important difference because crystalline cloud particles will have 
quite sharply peaked resonance features in the MIR (resolvable at $R\sim 
50$), while amorphous particles have much broader resonance features.

The assumption of amorphous cloud particles in exoplanets may be 
unphysical because the high temperatures under which cloud formation 
occurs should lead to condensation in crystalline form and/or annealing 
\citep{fabianjaeger2000,gail2001,harkerdesch2002,gail2004}. Note that the 
cloud is always in contact with high temperature regions close to the cloud 
base. Even particles which may form in higher and cooler layers above the 
cloud base should experience annealing due to mixing and/or settling to hotter 
regions of the atmosphere, if they did not condense in crystalline form in the 
first place. 

The fact that most silicates are present in amorphous form in the ISM is 
commonly attributed to the  ``amorphization'' of crystalline silicate grains by 
heavy ion bombardment, where the grains have been injected into the ISM by 
outflows of evolved stars in crystalline form \citep{kempervriend2004}. 
Because such processes are unlikely to occur in planetary atmospheres the 
assumption of crystalline particles may represent a better choice than 
amorphous particles.

Crystalline optical data were used for MgAl$_2$O$_4$, Mg$_2$SiO$_4$, Fe, and KCl, see Appendix \ref{sect:ref_cl_opa} for the 
references.

\subsubsection{Treatment of cloud self-feedback}
\label{cloud_instab}
The self-consistent coupling between the atmospheric temperature structure 
and the cloud model can in certain cases lead to oscillations and non-convergence
in atmospheric layers where the presence of the cloud heats the 
layer enough to evaporate the cloud. If this occurred in our calculations, then 
the cloud base location was moving significantly in the atmosphere. A similar 
behavior has been found for water cloud modeling in Y-dwarfs, using the same 
cloud model as one of the two which we adopted for the irradiated planets here
\citep[see][]{morleymarley2014}.

\rch{In our models, if the cloudy solution exhibited the unstable self-feedback behavior, we decreased the cloud density by multiplying it by 2/3 and waiting 100 iterations to check if the solution would settle into a stable state. This was repeated until a stable state was found.}

\rch{The motivation for this treatment is the following: A single temperature structure solution for an atmosphere with a cloud profile that leads to unstable cloud self-evaporation will on average have a lower cloud density. Physically this can be thought of as an average over the planetary surface where the clouds are in a steady state between condensation and evaporation. Alternatively, if there exist regions of rising and sinking parcels of gas a planet may well develop a patchy cloud pattern \citep{morleymarley2014}, such that our treatment may also be thought of as a an opacity-average over a patchy cloud model. In that sense our model is somewhat less sophisticated than the \citep{morleymarley2014} approach for self-luminous planets/brown dwarfs, where a single atmospheric temperature structure was calculated as well, but the radiative transport and emerging flux from the planet was calculated for the clear and cloudy atmospheric patches separately, with less flux emerging from the cloudy and more flux emerging from the clear parts of the atmosphere.
However, because for irradiated planets the majority of the flux does not stem from the cooling of the deep interior of the planet, but from the regions were the stellar flux is absorbed, the cloudy regions would have to re-radiate the same amount of energy as the cloud-free regions in the absence of thermal advection of energy. Thus our treatment may be more appropriate. However, from phase curves measurements and the corresponding day-nightside emission contrasts it is well known that the horizontal advection of thermal energy in irradiated planets is working quite effectively, indicating that the advection timescale becomes comparable to the radiative timescale. This advection seems to be most effective for cool, low-mass planets \citep{perez-becker2013,kammerknutson2015,komacekshowman2016}. Therefore, for cool, low-mass planets not all of the energy absorbed in a given region of the atmosphere is re-radiated immediately, which would in turn mean that the \citep{morleymarley2014} treatment may still be valid.}

\rch{We currently neglect the corresponding increase in the gas opacities due to the reduced cloud density because the condensates considered here do not significantly deplete the atmosphere's main opacity carriers (H$_2$O, CH$_4$, CO, CO$_2$, HCN, etc.): the atomic species such as Mg, Si and Al are all naturally less abundant when considering solar abundance ratios. The corresponding decrease in the gas abundance ratios are of the range of  $\sim$20~\% for water if silicate condensation takes place. The only exception is Na$_2$S and KCl which will deplete almost all Na and K from the gas phase if condensation occurs. However, because the unstable regions occur mostly at the location of the cloud bases, the evaporated Na and K gas is likely cold trapped to the cloud base regions such that the removal from the atmosphere's upper layers may still be valid. A more sophisticated treatment will be added in an upcoming version of the code.}

\rch{We publish the cloud density reduction factor for all cloudy atmosphere calculations.}

\subsection{C/O}
\label{sect:params_select:co}
The observational evidence of C/O$>$1 planets is debated 
\citep{madhusudhanharrington2011,crossfieldbarman2012,swainderoo2013,stevenson2014,lineknutson2014,kreidbergline2015,benneke2015} and there have been numerous 
studies trying to theoretically assess whether the formation of C/O$>$1, or ${\rm C/O}\rightarrow 1$, planets is possible \citep{obergmurray-clay2011,ali-dibmousis2014,thiabaudmarboeuf2014,hellingwoitke2014,marboeufthiabaud2014a,marboeufthiabaud2014b,madhusudhanamin2014,thiabaudmarboeuf2015,obergbergin2016,madhubitsch2016}, while one of the most recent works on this topic indicates that hot jupiters (which usually have masses $\lesssim 3 \mj$) and planets of lower mass may never 
have C/O$>$1 \citep{mordasinivanboekel2016}.

We want to note that the C/O ratio expected for a more massive planet with 
$M_{\rm Pl} > 3 \mj$, which can have a composition dominated by gas 
accretion, may never have a C/O value $>1$, but C/O values approaching 1 
are possible \citep[see, e.g.,][]{obergmurray-clay2011,ali-dibmousis2014}. The 
transition value from water to methane dominated spectra occurs for C/O 
values between $\sim$0.7 and $\sim$0.9 \citep{mollierevanboekel2015}, 
where the lower value for the transition is found in atmospheres which are 
cool enough to condense oxygen into silicates, increasing the gas phase C/O 
ratio \citep[also see][]{helling2014}.
Even cooler planets can exhibit methane features without the need for an 
elevated C/O ratio \citep{mollierevanboekel2015}.

In our calculations the fiducial composition of all planets is always a scaled 
solar composition with C/O~$\approx$~0.56 \citep{asplund2009}. For every 
planet we also consider models with twice as many or half as many O atoms, 
leading to C/O ratios of 0.28 and 1.12, respectively. While the value of 1.12 
may be slightly higher than can be reached from formation it obviously leads to the 
desired results, i.e. carbon-dominated atmospheres.

\subsection{TiO/VO opacities}
\label{sect:params_select:tiovo}
In cases where the target planets are hot enough for TiO and VO to exist in 
the gas phase at the terminator region we calculate additional models 
including TiO and VO opacities.

We want to repeat here that the existence of gaseous TiO and VO in planetary 
atmospheres is debated 
\citep{spiegel2009,showman2009,parmentier2013,knutson2010}. Recent 
evidence shows that this class of planets may exist nonetheless 
\citep{mancinisouthworth2013,haynesmandell2015,evanssing2016}, therefore we include this possibility here.

\subsection{Irradiation treatment}
\label{sect:irrad_treatment}
The atmosphere of the planet will be able to transport energy from the day to the nightside, thereby decreasing the flux of the planet measured during an occultation measurement. As shown in \citet{perez-becker2013,komacekshowman2016} this process depends on the equilibrium temperature of the planet: The hotter the equilibrium temperature, the weaker the redistribution becomes, meaning that radiative cooling increasingly dominates over advection.

To account for the different possibilities of energy transport we consider 3 different scenarios for our model calculations:

(i) globally averaged insolation, where the insolation flux is homogeneously spread over the full surface area of the planet, assuming the stellar radiation field impinges on the atmosphere isotropically.

(ii) day side averaged insolation, where the insolation flux is homogeneously spread over the dayside hemisphere of the planet, again assuming isotropic incidence.

(iii) case of no redistribution: for the very hot planets WASP-18b and Kepler-13Ab the brightness temperature of the dayside is higher than the temperature expected for both the dayside and the global average case \citep{nymeyerharrington2011,shporerorourke2014}. For these two planets we therefore calculated emission spectra by combining individual spectra for planetary annuli at angular distances $\theta$ between 0 and $\pi/2$ from the substellar point.
For every annulus we assumed that it has to reemit all the flux it received from the star, impinging at an angle $\theta$. Only the intensities of the rays headed into the direction of the observer were taken into account. This corresponds to the case where the energy advection by winds is fully neglected. 

\section{Results of the atmospheric calculations}
\label{sect:calc_res}
In this section we want to summarize the effect of various parameters on the 
resulting atmospheric structures and spectra, where more emphasis is put on 
the models including clouds. Note that we will publish the atmospheric 
structures, spectra and synthetic observations for all planets listed in Table 
\ref{tab:obs_cands}. For the sake of clarity, and in order to minimize 
redundancy, we concentrate on a selected subset of the candidates listed in 
Table \ref{tab:obs_cands} here, which we use to exemplary show the effects of 
various parameters.
 
\subsection{Atmospheric enrichment}
Variations of the atmospheric enrichment affect the resulting atmospheric 
structures and spectra in at least three different ways.

\begin{figure*}
\centering
\includegraphics[width=0.98\textwidth]{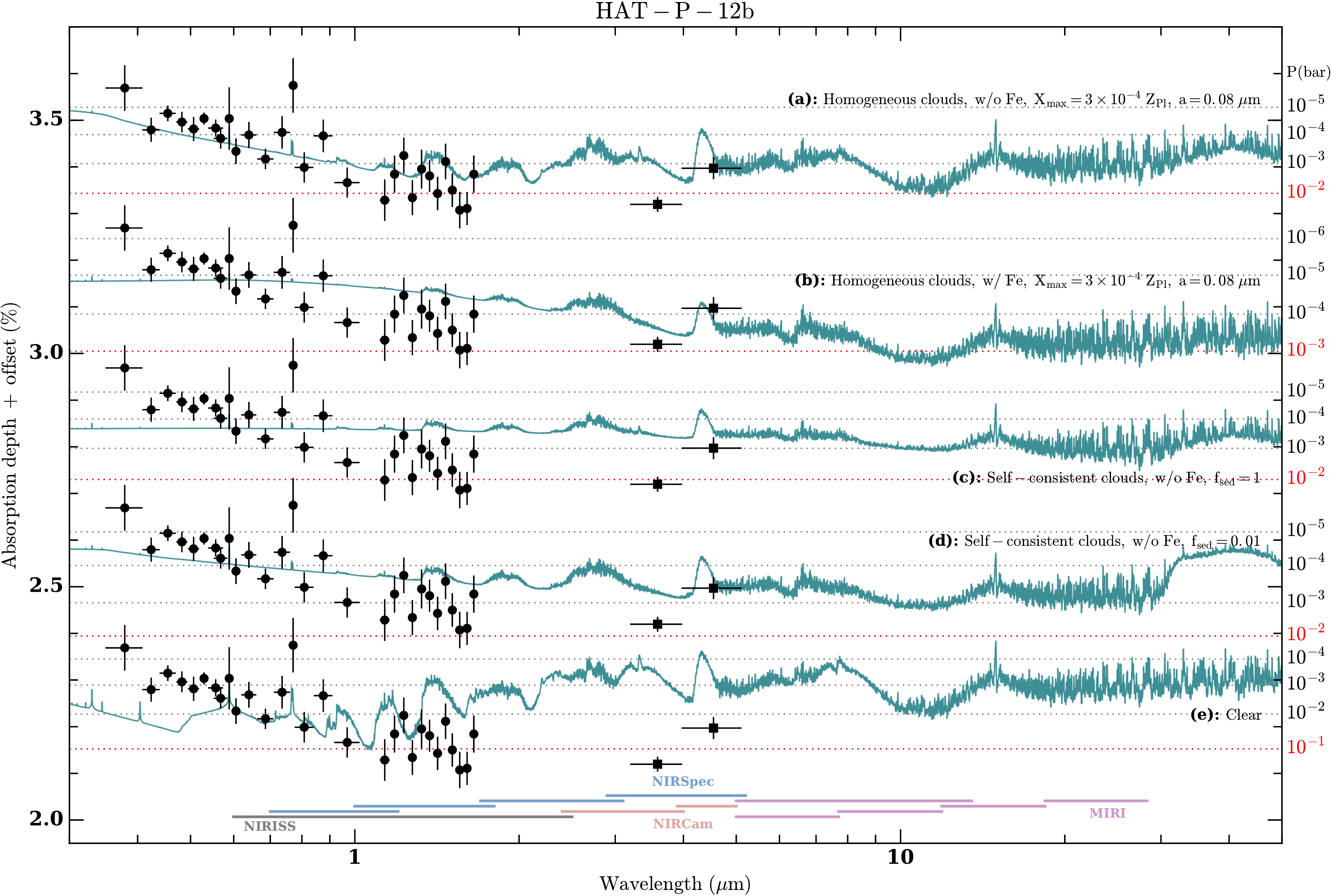}
\caption{Transmission spectra for the warm Saturn HAT-P-12b, along with the 
observational data taken from \citet{singfortney2015}. For clarity a vertical 
offset has been applied to the various models. \rch{The cloud species considered here are Na$_2$S and KCl only.}
From top to bottom the following cases are plotted: (a): homogeneous clouds, with a maximum cloud mass 
fraction of $X_{\rm max}=3\times10^{-4} \cdot Z_{\rm Pl}$ per species and a 
single cloud particle size of 0.08 $\mu$m. Iron clouds have been neglected; 
(b): like (a), but including iron clouds; (c): self-consistent clouds using 
the \citet{ackermanmarley2001} model with $f_{\rm sed}=1$, (d): like (c), but using $f_{\rm sed}=0.01$; (e): clear, fiducial atmospheric 
model. The colored bars at the 
bottom of the plot show the spectral range of the various \emph{JWST} 
instrument modes. The dotted horizontal lines denote the pressure levels 
being probed by the transit spectra with the pressure values indicated on the 
right of the plot.}
\label{fig:clouds_transm}
\end{figure*}

First, an increase (or decrease) of the enrichment will result in an increased 
(or decreased) total opacity within the atmosphere. This is because the main 
carriers of the atmospheric opacities are the metals, rather than H$_2$ and 
He. The effect of scaling the planetary enrichment on the atmospheric 
temperature structure and emission spectra has been studied in 
\citet{mollierevanboekel2015}, \rch{and we only provide a brief outline here}:
A higher enrichment moves the photosphere 
position to smaller pressures, leading to less pressure broadening of lines \rch{and a decreased strength of the CIA opacity because the strength of both these opacity sources scales linearly with pressure. Hence the opacity in the atmospheric windows decreases.} This exposes \rch{deeper, hotter layers in the windows} and leads to a larger contrast between emission minima and maxima in spectra. This effect, neglecting metallicity-dependent chemistry, is (inversely) degenerate with varying the 
planetary surface gravity as it holds that $d\tau_\nu=(\kappa_\nu/g)dP$, where 
$\tau_\nu$ is the optical depth, $\kappa_\nu$ the opacity, $g$ the surface 
gravity and $P$ the pressure.

Second, atmospheric transmission spectra are affected by scaling the 
metallicity in 2 ways: Increasing the metallicity and therefore the total opacity 
will result in an increased transit radius, while a significant increase in 
metallicity and the resulting increase of the atmospheric mean molecular 
weight will weaken the signal amplitude between maxima and minima in the 
transmission spectrum because $R(\lambda) \propto [k_{\rm B}T/(\mu g)]
\cdot{\rm log}(\kappa_\nu)$ \citep{etangspont2008}, where $R(\lambda)$ is 
the planetary transit radius, $k_{\rm B}$ the Boltzmann constant, $T$ the 
atmospheric temperature, $\mu$ the atmospheric mean molecular weight and 
$g$ the planetary surface gravity.

Finally, an increased enrichment will affect the atmospheric abundances 
because of the metallicity-dependent chemistry: the CO$_2$ abundance, e.g., 
is a strong function of metallicity \citep[see, e.g.,][and the references therein]
{moses2013}.

\subsection{Clouds}
\label{sect:clouds_spec}
The various cloud models investigated in this study are summarized in Section 
\ref{sect:params_select:clouds}, here we concentrate on the spectral 
characteristics of some of these cloud models. In Figure 
\ref{fig:clouds_transm} we show transmission spectra resulting from self-consistent atmospheric structure calculations of the planet HAT-P-12b, 
together with the observational data of the \emph{HST} and \emph{Spitzer} 
telescope \citep{singfortney2015}. \rch{We look at the cases including only Na$_2$S and KCl clouds here.} For these calculations a planet-wide 
averaged insolation was assumed, because it was found that 3D GCM 
calculations lead to similar transmission spectra as the planet-wide averaged 
insolation 1D modes \citep[but exceptions may exist, for more details see][]
{fortneyshabram2010}.
Effects of patchy clouds, which may mimic the signal of high mean molecular 
weight atmospheres \citep{lineparmentier2016}, cannot be reproduced with 
this approach.

The model spectra plotted in Figure \ref{fig:clouds_transm} include the small 
and \rch{larger} cloud particle ($f_{\rm sed}$=\rch{0.01} and \rch{1}, respectively) self-consistent 
clouds following the \citet{ackermanmarley2001} cloud model, as well as the 
parametrized homogeneous clouds with small particles of size 0.08 $\mu$m. Models are shown with 
and without the consideration of iron clouds. Finally, our fiducial, cloud free model is shown as well.
Note that we also draw horizontal lines in the plot which indicate the pressure being probed 
by the various models, corresponding to the (wavelength-dependent) effective radius.
The optimal y-offset value of the spectra with respect to the data was found by $\chi^2$~minimization.

Before investigating the different cloudy models we want to note that the clear 
atmosphere (Model \rch{(e)} in Figure \ref{fig:clouds_transm}) obviously represents 
a bad fit to the data: a simultaneous fit of the Rayleigh like signature of the 
optical and NIR \emph{HST} data and the \emph{Spitzer} photometry at IR 
wavelengths is not possible. Note that the \emph{HST} \emph{STIS} and 
\emph{Spitzer} data are both crucial for this claim of ``cloudiness'' because the 
measurement of a Rayleigh \rch{signal} in the optical alone is not sufficient as it 
could be simply caused by H$_2$ and He. Only a spectral slope less negative than 
$-4$ in the optical may allow to infer the presence of large particle ($a\gtrsim
\lambda/2\pi$) clouds from the spectrum alone, but this requires an accurate 
estimate of the atmospheric scale height \citep[also see][]{heng2016}.

Studying the cloudy results in Figure \ref{fig:clouds_transm} one sees that 
only the parametrized homogeneous clouds with a particle size of 0.08~$\mu
$m (model (a) in Figure \ref{fig:clouds_transm}) are able to produce 
Rayleigh \rch{scattering} ranging from the optical ($\sim 0.4$~$\mu$m) to the NIR as 
probed by the \emph{HST} data.
Further, this is only possible if iron clouds are neglected: Due to the 
high absorptivity of iron in the optical the Fe clouds clearly 
decrease the spectral slope such that it is less strong than expected for pure 
Rayleigh scattering (see Model \rch{(b)} in Figure \ref{fig:clouds_transm}).

\begin{figure}[t]
\centering
\includegraphics[width=0.495\textwidth]{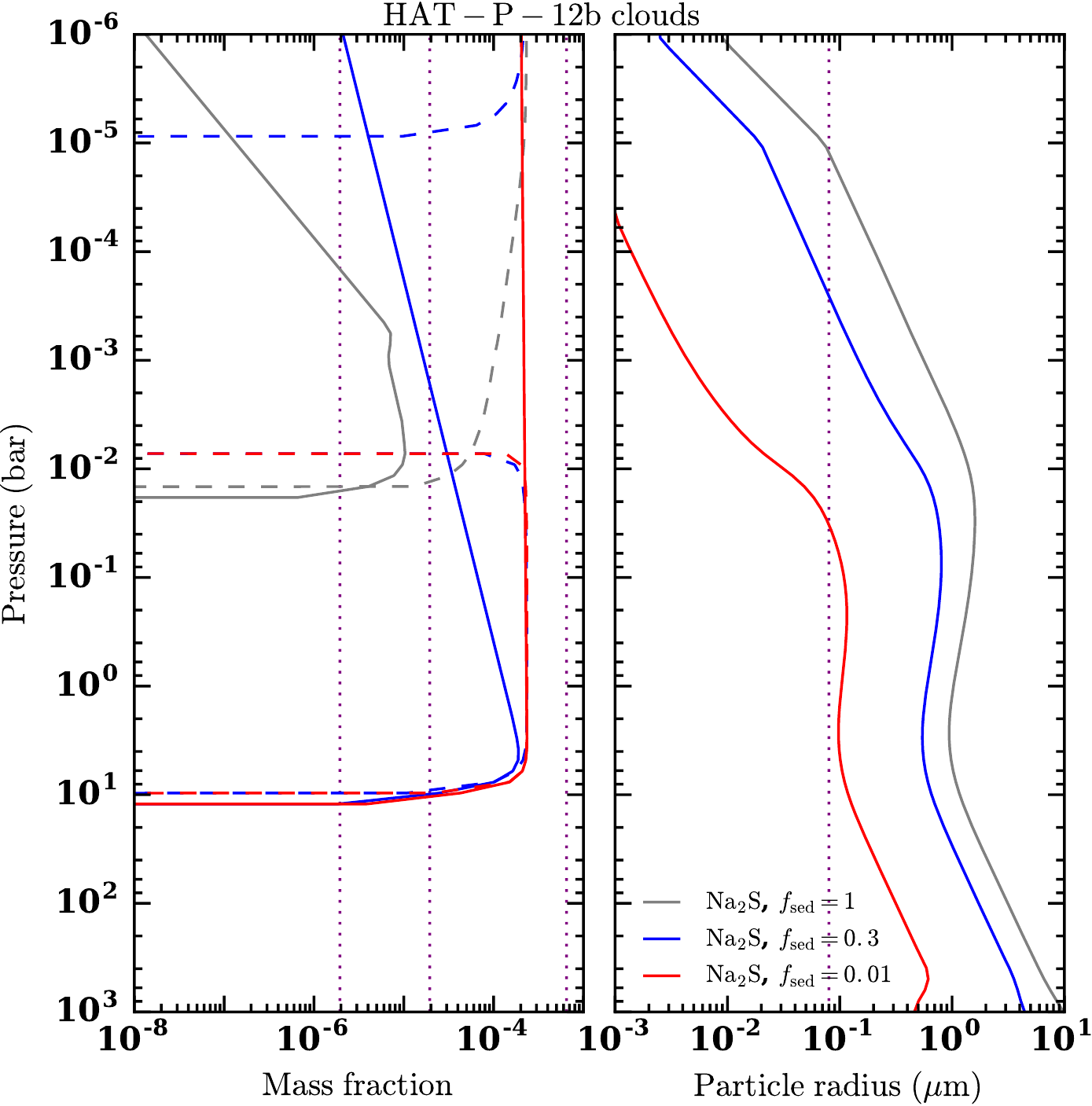}
\caption{{\it Left panel}: \rch{the solid lines denote the Na$_2$S cloud mass fractions as a 
function of pressure for the $f_{\rm sed}=0.01$, 0.3 and 1 models, shown in red, blue and gray, respectively.} The dashed lines denote the mass fractions derived from 
equilibrium chemistry, i.e. in the absence of mixing and settling. For a 
comparison, the four vertical lines denote the 3 different $X_{\rm max}$ values 
used in the homogeneous cloud models. {\it Right panel}: Mean cloud particle 
radii derived for the \rch{$f_{\rm sed}=0.01$, 0.3 and 1} cloud models.
Again for comparison the vertical dotted line denotes the 
particle size $a=0.08$~$\mu$m adopted for the homogeneous cloud models.}
\label{fig:hatp12_cloud_dens}
\end{figure}

\rch{The results for the \citet{ackermanmarley2001} clouds are shown in the models \rch{(c)} and \rch{(d)} 
in Figure \ref{fig:clouds_transm} for $f_{\rm sed}$=1 and \rch{0.01}, respectively. Model (c) a produces flat slope in the optical and NIR and seems to mute the molecular features too strongly when compared to the data.
For the self-consistent cloud with a small $f_{\rm sed}$=0.01 value (model d) we find that the slope in the optical is quite steep already, approaching a Rayleigh scattering slope. However, although the average particle size is well below 0.08~$\mu$m the slope is less steep than in the mono-disperse particle model (a) because the largest particles within the distribution dominate the opacity \citep{wakefordsing2015}. We thus do not find a good fit for HAT-P-12b when using the \citet{ackermanmarley2001} model. The broad absorption feature starting at 30~$\mu$m in the transmission spectrum of model (d) is the Na$_2$S resonance feature.}

\rch{We show the mean particle size obtained for the $f_{\rm sed}
$=0.01, 0.3 and 1 models in the right panel of Figure \ref{fig:hatp12_cloud_dens}, along with 
the cloud mass fractions in the left panel. Only Na$_2$S condensed for the models presented here, because the atmosphere was too hot for KCl condensation to occur.}

Note that the cloud mass fraction derived from our cloud model 
implementation starts already one layer below the layer where equilibrium 
chemistry first predicts condensation. This is due to the fact that the cloud source 
term in our implementation of the \citet{ackermanmarley2001} model is 
proportional to $d X^{\rm equ}_{\rm c}/dz$, where $X^{\rm equ}$ is the mass 
fraction of the condensate species derived from equilibrium chemistry. This 
derivative, evaluated between the two layers, leaves a non-zero cloud mass 
fraction at the lower layer when solved on a high resolution grid between the 
layers and then interpolated back to the coarse resolution. The advantage of 
our implementation is that no knowledge on the saturation pressures of given 
condensates is required, and that one can simply use the output abundances 
of a Gibbs minimizer (see Appendix \ref{sect:app_cloud_mod} for more 
information). The true location of first condensation is located somewhere 
between the two layers. Due to the good agreement when comparing to the 
example cases given in \citet{ackermanmarley2001} we decided to keep the 
current treatment.\footnote{For the comparison study we adopted the same 
mixing length choice as in \citet{ackermanmarley2001}.}

\rch{Due to the large size of the cloud particles shown for the $f_{\rm sed}=1$ model in Figure
\ref{fig:hatp12_cloud_dens} cloud resonance features in the MIR are hard to see in Figure \ref{fig:clouds_transm} (model c): Only small enough grains exhibit resonance features, while 
increasingly larger grains have flatter, more grayish opacities. Consequently, the Na$_2$S feature at 30~$\mu$m can be seen more prominently for model (d), as this model results in smaller particle sizes.}

\rch{To summarize, within our work, the \citet{ackermanmarley2001} results correspond to clouds with broader particle size distributions which tend to produce transmission spectra that are either flat or less strongly sloped than 
expected from small particle Rayleigh scattering. For the smallest $f_{\rm sed}$ values we succeed in getting quite steep scattering slopes, but at the same time the clouds are quite optically thick, muting the molecular features more strongly.}
We want to point out that this does not mean that the 
\citep{ackermanmarley2001} model is not useful for fitting cloudy planetary 
spectra, but it is \rch{potentially} more useful for transmission spectra that show a flat and gray 
cloud signature in the transmission spectrum, as is seen for GJ~1214b 
\citep[see][and our results for GJ~1214b using cloud model 4 in Section \ref{sect:transm_obs}.]
{morleyfortney2015}

Finally, some of the spectra we show in Figure \ref{fig:clouds_transm} may fit 
the transmission results quite well. \rch{We want to remind the reader that we used the same cloud setups for all candidate planets presented in this paper, without trying to find the true best fit parameters according to our model. Therefore, a dedicated study for the individual planets may result in even better estimates of the atmospheric parameters.}
Furthermore, it is not correct to assume that the homogeneous clouds 
used for model (a) represent a good description of the cloud mass fraction and 
particle sizes throughout the whole atmosphere. As can be read off in Figure 
\ref{fig:clouds_transm}, the maximum pressure being sensed by transmission 
in model (a) is around \rch{10~mbar}. Therefore, an equally good fit may be obtained 
using a cloud model which truncates the cloud at \rch{$P>10$~mbar} and sets the cloud 
density to zero at larger pressures. This may leave the transmission spectrum 
unchanged but will strongly affect the planet's emission spectrum: as 
mentioned previously, it is well known that the emission spectra probe higher 
pressures than transmission spectra. This is due to the different trajectories of 
the light rays probing the atmosphere vertically in emission vs. the grazing 
geometry of light rays during transmission spectra. The vertical vs. the grazing 
optical depth at the photosphere of the planet may be different by a factor
$\tau_{\rm trans}/\tau_{\rm emis} \sim \sqrt{R_{\rm Pl}/H}$, see 
\citet{fortney2005}, which easily reaches a factor of 100 or larger.
For instance, the retrieval analysis of HD~189733b requires clouds to fit the 
transmission spectrum, while the emission spectrum is fit well with a clear 
atmosphere \citep{barstow2014} (a similar result could be obtained when considering a clear dayside and cloudy terminator regions, however).

\begin{figure}[t]
\centering
\includegraphics[width=0.495\textwidth]{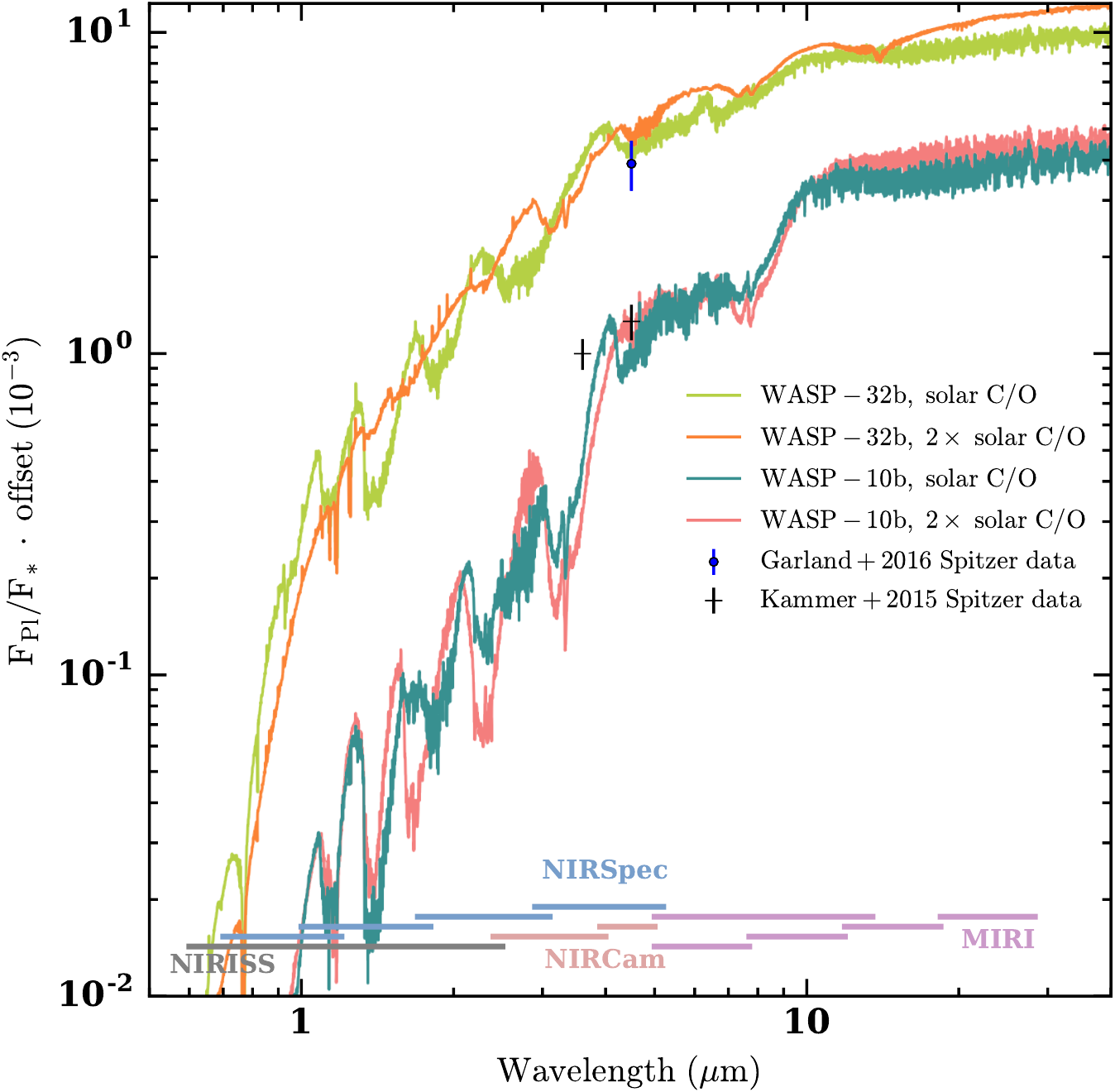}
\caption{Synthetic emission spectra of the planets WASP-32b and WASP-10b 
for the clear fiducial models with solar C/O ratios and for the cases with C/O 
ratios twice as large as solar. We also plot existing \emph{Spitzer} photometry 
for the targets by \citet{garlandharrington2016} (WASP-32b) and 
\citet{kammerknutson2015} (WASP-10b). For both planets a dayside-averaged insolation was assumed. For clarity both the synthetic spectra and 
data of WASP-32b have been multiplied by an offset factor of 3.}
\label{fig:COemis}
\end{figure}

Due to the assumption of homogeneous clouds we obtain a Bond albedo 
of \rch{17~\%} for model (a) in Figure \ref{fig:clouds_transm}. 
Model (b), which has the same $X_{\rm max}$ value as model (a), only has an albedo of \rch{5~\%}, because the iron 
cloud particles absorb light effectively. 
For the self-consistent clouds we obtain \rch{10~\%} for Model (c) and \rch{8~\%} for 
Models (d). The clear model (e) has an albedo of 3~\%.
If clouds were truncated below the pressures probed by transmission the albedos for the models would be lower.

In conclusion it may therefore well be the case that the transmission spectrum 
of a planet is fit well by a cloudy atmosphere, while the planet's emission 
spectrum is described well by the corresponding clear atmosphere. For HAT-
P-12b this assessment is impossible because the \emph{Spitzer} eclipse 
photometry for the dayside emission by \citet{todorovdeming2013} only gives 
upper limits at 3.6 and 4.5~$\mu$m. These limits are consistent with all model 
calculations we carried out for this planet for a planet-wide averaged 
insolation. The dayside averaged insolation case is excluded, because all models have larger fluxes than allowed by the upper limits.

\subsection{C/O}
\rch{The importance of the C/O~ratio for atmospheric chemistry and the effects arising from varying this parameter have been described in, e.g., \citet{seager2005,kopparapu2012,madhusudhan2012,moses2013}}.
Additionally, the dependence of properties of planetary atmospheres upon variation of the C/O ratio has been extensively and systematically studied in our previous study \citet{mollierevanboekel2015}, such that we only give a short summary here.

In Figure \ref{fig:COemis} we show the emission spectra calculated for the 
warm Jupiter WASP-10b ($T_{\rm equ}=972$~K) and the hot Jupiter 
WASP-32b ($T_{\rm equ}=1560$~K), along with the \emph{Spitzer} eclipse 
measurements by \citet{kammerknutson2015} and 
\citet{garlandharrington2016}, respectively. Note that both the synthetic 
spectra and data of WASP-32b have been multiplied by an offset factor of 3 in 
order to minimize the overlap between the spectra of WASP-10b and 
WASP-32b. We plot the spectra resulting from a dayside averaged insolation 
for both planets.

For the hotter planet, WASP-32b, the spectra exhibit a clear dichotomy, with 
the solar C/O spectrum (C/O$_\odot\sim0.56$) dominated by water absorption 
and the spectrum with twice the solar C/O value (1.12) dominated by methane 
absorption. 

For the cooler planet, WASP-10b, the spectrum of the solar C/O case shows 
both water and methane absorption (see the telltale methane feature at 3.3 $
\mu$m), while the C/O=1.12 case shows methane absorption but no water 
absorption. The reason that lower temperature atmospheres can show both 
water and methane absorption at the same time, regardless of the C/O ratio, 
can be understood by considering the net chemical equation
$\ce{CH_4 + H_2O <=> CO + 3 H_2}$, where the left-pointing direction is 
favored for low temperatures ($\lesssim 1000$~K) and high pressures \rch{\citep[see, e.g.,][]{lodders2010}}. Note 
that depending on the vigor of vertical mixing the transition temperature below 
which the left-pointing direction is favored in the atmospheres of planets may 
be as low as 500~K due to chemical quenching \citep{zahnlemarley2014}. We 
do not model such non-equilibrium chemistry effects in our calculations. 
However, the importance of non-equilibrium chemistry strongly depends on 
the values of the mixing strength which is related to the planetary surface 
gravity. Both planets shown here have rather high surface gravities, which 
tends to decrease the mixing strength \citep{zahnlemarley2014} and shifts the 
location of the photosphere to higher pressures where the left-pointing direction of the above reaction equation is favored even more.

\subsection{TiO/VO opacities}
\label{sect:tiovo_behavior}
In the cases where equilibrium chemistry allows for TiO and VO to exist in the 
gas phase, and for the calculations for which we specifically include TiO and 
VO opacities, we find that the converged atmospheric solutions exhibit 
inversions. For cases where the atmospheres are cool enough such that Ti 
and V have condensed out of the gas phase we find that the results are 
identical to our clear, fiducial calculations, which do not include TiO/VO 
opacities.

Atmospheres with TiO and VO inversions show emission spectra which are 
more isothermal than the corresponding fiducial cases, i.e. the SED more 
closely resembles a blackbody, because the inversion decreases the overall 
temperature variation in the {\it photospheric layers} of the atmospheres. \rch{Note that this does not mean that the atmospheres attain a globally more isothermal state, we still find strong inversions if the insolation is strong and the TiO/VO abundances are high enough. The decreasing temperature variability merely holds for the photospheric region, not for the whole atmosphere: Inversions form if the opacity of the atmosphere in the visual wavelengths is larger than in the IR wavelengths. When entering the atmosphere from the top the optical depth for the stellar light reaches unity before the location of the planetary photosphere is reached. Therefore the higher atmospheric layers in which a non-negligible amount of the stellar light is absorbed need to heat up significantly in order to reach radiative equilibrium (i.e. absorbed energy equals radiated energy). These layers will cause the formation of emission features. On the other hand, the photosphere represents the region where the planet's atmosphere radiates most of its flux to space, because here the IR optical depth reaches unity. This region is below the inversion region. Below the photosphere the atmospheric temperature will increase monotonously. Consequently, the photosphere is bracketed by a region where the temperature decreases as one approaches the photosphere coming from the inversion above, and a region where the temperature increases again when moving on to larger pressures. Hence the total temperate variation across the photospheric region, which is a region in which the atmospheric temperature gradient transitions from being negative to being positive, is small. Therefore the spectral energy distribution escaping from this region is closer to an isothermal blackbody than in an atmosphere without an inversion.}

In transmission these atmospheres exhibit TiO/VO resonance features in the optical and NIR, which are well known from theoretical calculations of atmospheric spectra \citep[see, e.g.][]{fortney2008,fortneyshabram2010} but have not yet been conclusively detected in observations.

If we include TiO/VO opacities all atmospheres with equilibrium temperatures higher than 1500 K show 
inversions in their atmospheres for the dayside averaged insolation 
calculations. In planet-wide averaged insolation calculations these planets 
showed inversions as well but the planets with equilibrium temperatures below 
1750 K had inversions which were quite high in the atmospheres, such that 
either none or only weak emission features were seen. The transmission 
signatures of TiO/VO were seen in all cases which exhibited an inversion.

\begin{table}[t]
\begin{center}
\begin{tabular}{lccc}
\hline \hline 
 & \emph{NIRISS SOSS I}  &  \emph{NIRSpec G395M} &  \emph{MIRI LRS} \\
\hline
$\lambda$ range    & 0.8-2.8~$\mu$m             &  2.9-5.0~$\mu$m       & 
5.5-13.5~$\mu$m  \\
QE  & 0.8                      & 0.8                 & 0.6                \\
FWC  & 77\,000 e-               & 77\,000 e-          & 250\,000 e-      \\
$N_{\rm read}$       & 23 e-                    & 18 e-                & 14 e-            \\
DC        & 0.02 e- s$^{-1}$         & 0.01 e- s$^{-1}$     & 0.2 e- s$^{-1}$   \\
$T_{\rm tot}$ &  0.15              & 0.54                & 0.35             \\
$N_{\rm floor}$ &  20~ppm              &  75~ppm             &  40~ppm        \\
\hline \hline
\end{tabular}
\end{center}
\caption{\label{tab:JWST_parameters} Instrument parameters values used for 
the synthetic \emph{JWST} observations. The collecting area of \emph{JWST} 
is assumed to be 24\,m$^2$, with a ``warm'' mirror temperature of 35\,K. The 
abbreviations in the first column stand for the {\it quantum efficiency} (QE), the 
{\it full well capacity} (FWC), the {\it readout noise} ($N_{\rm read}$), the {\it 
dark current} (DC), the {\it total system transmission} ($T_{\rm tot}$) and the 
{\it systematics noise floor} $N_{\rm floor}$.}
\end{table}

\begin{figure*}
\centering
\includegraphics[width=0.485\textwidth]{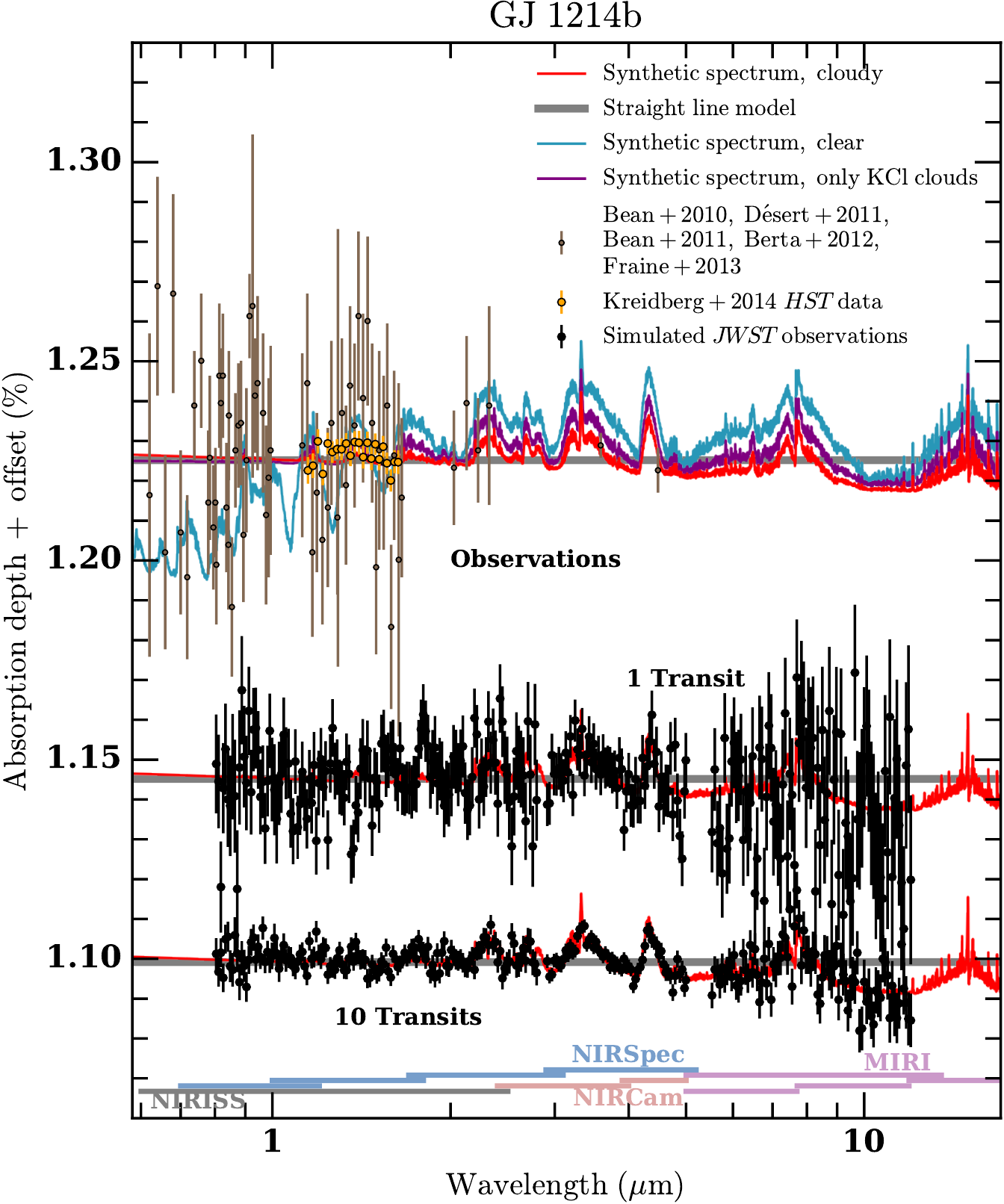}
\raisebox{0.01\height}{\includegraphics[width=0.475\textwidth]{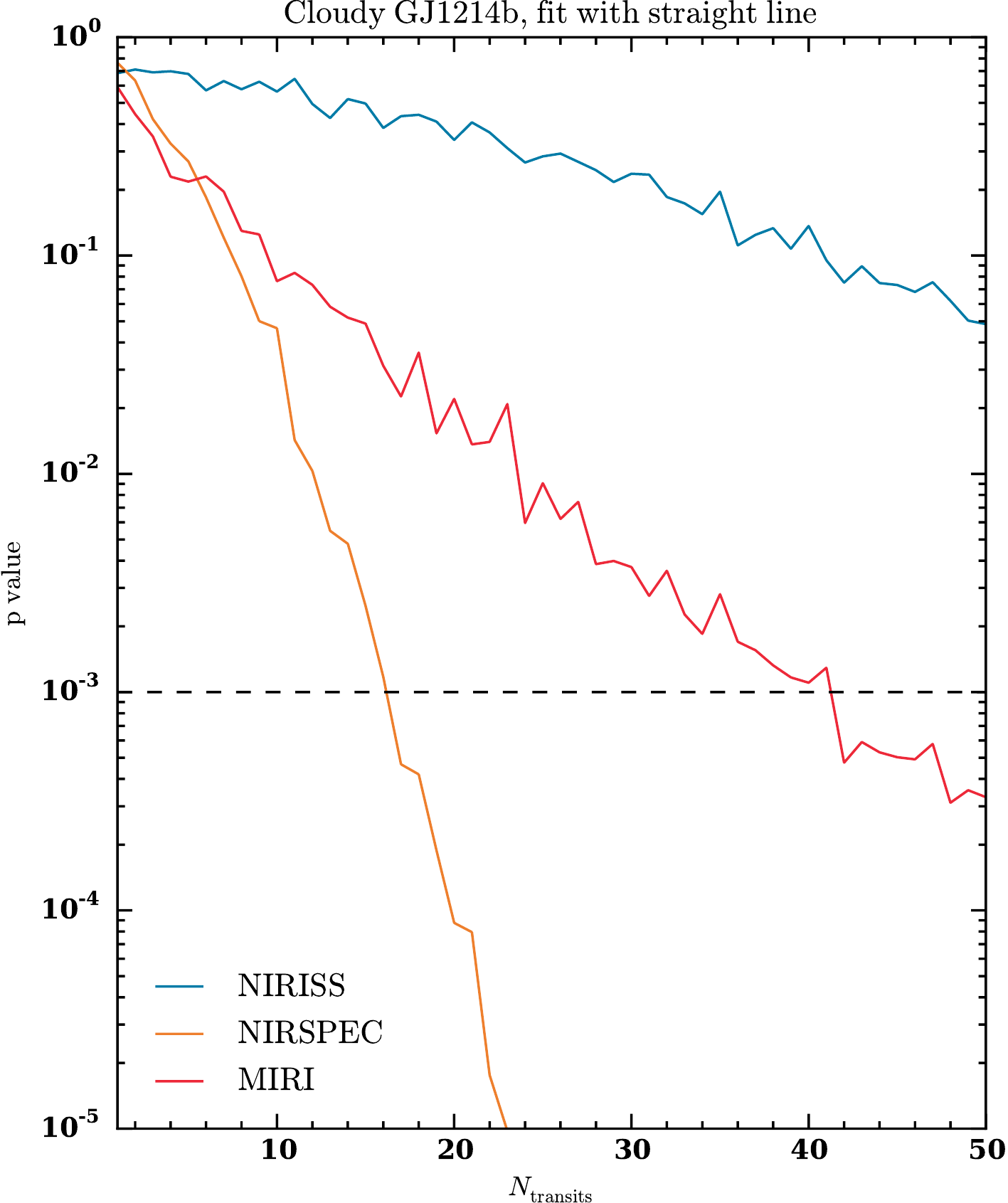}}
\caption{{\it Left panel:} synthetic transit spectra, observations, and synthetic observations for the planet GJ~1214b.
The orange points denote the observational data by \citet{kreidbergbean2014}, while brown points denote the observational data by \citet{beanmiller2010,desertbean2011,beandesert2011,bertacharbonneau2012,frainedeming2013}. Synthetic spectra for the cloudy $f_{\rm sed}=0.3$ model (Model 3 in Table \ref{tab:cloud_models}) are shown as red or purple solid lines for the case including Na$_2$S+KCl clouds or KCl clouds only. The clear model is shown as a teal line. A straight line model is shown as a thick gray solid line. The black dots show the synthetic observations 
derived for 1 (top) and 10 (bottom) transits, re-binned to a resolution of 50. Vertical offsets have been applied for the sake of clarity. {\it Right panel:} $p$~values of the Kolmogorov-Smirnov test for the residuals between the synthetic observation of the Na$_2$S+KCl cloud model and the straight line model fitted to the these observations. The $p$~value is shown as a function of $N_{\rm transit}$ for the 3~different instruments of Table~\ref{tab:JWST_parameters}. For every (instrument, $N_{\rm transit}$) setup a new straight line model is fitted to the observations. The black dashed line denotes our threshold value of $10^{-3}$.}
\label{fig:gj1214_transm}
\end{figure*}

\section{Simulated observations}
\label{sect:sim_obs}
In this section we will show the characteristics of the simulated observations 
carried out for all targets defined in Table \ref{tab:obs_cands}. Similar to 
Section \ref{sect:calc_res} we will concentrate on a few, exemplary objects. 
Here we will investigate how the simulated observations look like as a 
function of the number of transits/eclipses, and which wavelength ranges have the most diagnostic power for characterizing the planets.

The instrument parameters adopted to describe the performance of 
\emph{JWST} are listed in Table \ref{tab:JWST_parameters}.
The values for the full well capacity and readout noise of the \emph{NIRSpec} instrument were taken from \citet{ferruitbirkmann2014}, and we adopted the same full well capacity for \emph{NIRISS}, due to the similarity of the detectors. The noise floor for \emph{NIRSpec} is expected to be below 100~ppm \citep{ferruitbirkmann2014}, and we adopt a value of 75~ppm here. Following \citet{rocchetto2016} one may assume a noise floor value of 20 ppm for \emph{NIRISS}. Further, we set the \emph{MIRI} noise floor value to 40 ppm, because the values adopted in the existing literature range from 30 to 50 \citep[see][respectively]{beichmanbenneke2014,greeneline2016}. The remaining instrument characteristics for \emph{MIRI} were taken from \citep{resslersukhatme2015}.

For every planet and every instrument we publish synthetic observations corresponding to a single transit or eclipse measurement. Note that for every instrument a separate observation will have to be carried out. The data is given at the instruments' intrinsic resolutions. In regions where the wavelength binning of \emph{petitCODE} was coarser than the intrinsic resolution we rebinned the spectra of \emph{petitCODE} to this higher resolution.
Lower resolution data may be obtained by rebinning the observations and propagating the errors during the process. We also publish the model spectra without observational noise along with the single observation errors, such that multiple transit/eclipse observations can be obtained by sampling the noiseless spectra using errors normalized with $N_{\rm transit}^{1/2}$.

\subsection{Transmission spectroscopy}
\label{sect:transm_obs}

In this section we chose two different kinds of planets to show examples of our 
synthetic observations: super-Earth planets with flat transmission spectra, as 
well as hot jupiters with steep Rayleigh signals. The target chosen 
here for the super-Earths is GJ~1214b, \rch{while for the hot jupiters we 
investigate TrES-4b}.

\subsubsection{The case of extremely cloudy super-Earths: GJ~1214b}
\label{subsubsect:gj1214b}
The observational data for GJ~1214b, as well as synthetic spectra and 
observations are shown in the left panel of Figure \ref{fig:gj1214_transm}. For clarity the synthetic observations have been re-binned to a resolution of 50. Note that the noise of 
the measurements increases with wavelength as less light is coming from the 
star at longer wavelengths. In addition to the cloudy $f_{\rm sed}=0.3$ model (Model 3 in Table \ref{tab:cloud_models}) shown in the plot we also show a clear atmosphere for comparison. For this planet cloud models 3 and 4 converged with the cloud feedback included. We therefore present self-consistent calculations for cloud model 3 for GJ~1214b.

It is evident that the clear spectrum is inconsistent 
with the \emph{HST} data by \citet{kreidbergbean2014}, whereas cloudy 
models provide a better fit. The need for clouds has been studied in detail in 
\citet{morleyfortney2013,kreidbergbean2014,morleyfortney2015}, where 
\citet{morleyfortney2013,morleyfortney2015} found that high atmospheric enrichments are necessary to fit the data. They also put forward the possibility 
that the flat transmission spectrum of GJ~1214b could be caused by 
hydrocarbon hazes, and suggested pathways of how to distinguish between 
mineral clouds and hydrocarbon hazes using emission spectroscopy or by 
analyzing the reflected light from these planets.

Our cloudy spectrum is mostly flat from the optical to the NIR, but some molecular features can be made out clearly especially in MIR region, including the methane features at 2.3, 3.2 and 7.5~$\mu$m and the CO$_2$ features at 2.7 and 4.3 and 15~$\mu$m. The CO$_2$ feature at 15~$\mu$m is 
not within the spectral range of the \emph{MIRI LRS} instrument. Due to the high metallicity CO$_2$ is the most spectrally active carbon and oxygen bearing molecule and more abundant than CH$_4$, CO, H$_2$O at the pressures being probed by the transmission spectrum. For a cloudy, highly enriched 
atmosphere as presented here we therefore predict the existence of CO$_2$ and CH$_4$ features in the otherwise flat transmission spectrum.

Because GJ~1214b is the coolest planet considered in our sample, we only include KCl and Na$_2$S clouds in its atmosphere (see Section \ref{sect:params_select:clouds}). However, even Na$_2$S clouds may form too deeply in this atmosphere such that they can not be mixed up into the the region probed in transmission \citet{morleyfortney2013,charnaymeadows2015}.
Both approaches, i.e. excluding Na$_2$S or including it, have been studied in the literature \citep{morleyfortney2013,charnaymeadows2015b,morleyfortney2015}. We thus also show a comparison to a model only including KCl clouds in Figure \ref{fig:gj1214_transm}. One sees that as the cloud opacity decreases the molecular features can be seen more clearly.

The highest quality spectrum currently available for GJ~1214b is consistent with a straight line \citep{kreidbergbean2014}. We therefore want to assess how well \emph{JWST} observations could distinguish our high metallicity cloudy model from a flat featureless spectrum, as a function of the number of transits observed. As an example we show a comparison model, a straight line spectrum, as a gray line in Figure \ref{fig:gj1214_transm}. The transit radius of the straight line model was chosen by fitting a synthetic single transit observation of the KCl+Na$_2$S cloud model in all instruments with a straight line by means of $\chi^2$~minimization.

First we will test which instrument, i.e. wavelength range, is best suited for the task and how many transits are needed to conclusively rule out the straight line case. In order to avoid ruling out a straight line scenario because of an offset of the global (fitted for all 3 instruments) straight line model to a single instrument spectrum we fitted straight line models to the synthetic observation within each instrument separately.
For this we generated synthetic observations $T_\lambda({\rm instrument},N_{\rm transit})$, where the $T_\lambda({\rm instrument})$ denotes the observed wavelength-dependent transmission using one of the instruments listed in Table \ref{tab:JWST_parameters} and $N_{\rm transit}$ is the number of transits accumulated to obtain the observation. For every (instrument,$N_{\rm transit}$) pair we then fitted a straight line to the observations and calculated the residuals of the straight line to the cloudy observation, taking into account the appropriate errors when stacking $N_{\rm transit}$ transits in the instrument of interest. The residuals were then compared to a Gaussian normal distribution using a Kolmogorov-Smirnov test.\footnote{We used the kstest() function of the Scipy library for this task, see \url{http://docs.scipy.org}.} For ruling out the straight line model, given our observation of the cloudy model, we adopted a conservative threshold $p$~value of $10^{-3}$. This means that the probability of observing a straight line model and finding the above distribution of residuals, or one that is even less consistent with a normal distribution, is $10^{-3}$.
Alternatively to fitting a straight line model to the synthetic observations it is also possible to adopt an arbitrary offset between the model and the observations and then shift the distribution of residuals between the straight line model and the cloudy model such that is has a mean value of zero. For this case we obtained identical results.

We show the resulting $p$~values as a function of $N_{\rm transit}$ in the right panel of Figure \ref{fig:gj1214_transm} for the 3 different instruments. To minimize the Monte~Carlo noise resulting from the generation of the synthetic observations we took the median $p$~value of 100 realizations for every (instrument,$N_{\rm transit}$) point. It is found that \emph{NIRISS} will not be able to rule out a straight line spectrum even if 50~spectra are stacked. This is due to the fact that the cloudy model is quite flat in this wavelength region. With \emph{NIRSpec}, on the other hand, the distinction may be possible by stacking 16~observations. For \emph{MIRI} a refutal of the straight line model is possible after $\sim$40 transits.
Thus, using our conservative $p$~value threshold, it seems quite hard to refute the straight line case, although the \emph{NIRSpec} observations look different from a straight line when inspected by eye already after less than 10 transits in Figure \ref{fig:gj1214_transm}. Therefore, if one carries out the same test once more, but compares the observations with the cloudy model itself, then one finds $p$~values with a median of $1/2$, independent of the number of transits. Hence one may say that a cloudy model is more likely to describe the data than the straight line model, already after less than $\sim$10 transits, if one uses the \emph{NIRSpec} band and carries out a retrieval analyses. Note, however, that the $p$~value is subjected to statistical noise due to the limited number of spectral points.

The reason for the Kolmogorov-Smirnov test to require a quite large number of transits for the distinguishability analysis presented here is that the triangularly shaped molecular features will lead to quite symmetric residual distributions when compared to the straight line model. In this sense it becomes harder for the Kolmogorov-Smirnov test to tell the difference between the resulting residual distribution and a Gauss distribution, because this test is insensitive to the wavelength correlation of the residuals.
For a conclusive statement, rather than an upper limit, regarding the number of transits needed for constraining GJ~1214b's atmosphere one therefore needs more sophisticated statistical tools, such as retrieval analyses. However, the Kolmogorov-Smirnov test may still be used to assess which instrument, and thus wavelength range, may be best suited to distinguish the cloud GJ~1214b observation from a straight line and our analysis indicates that \emph{NIRSpec} will be best suited for this task, followed by \emph{MIRI}. With \emph{NIRISS} such a distinction will be the most difficult.

\begin{figure*}
\centering
\includegraphics[width=0.484\textwidth]{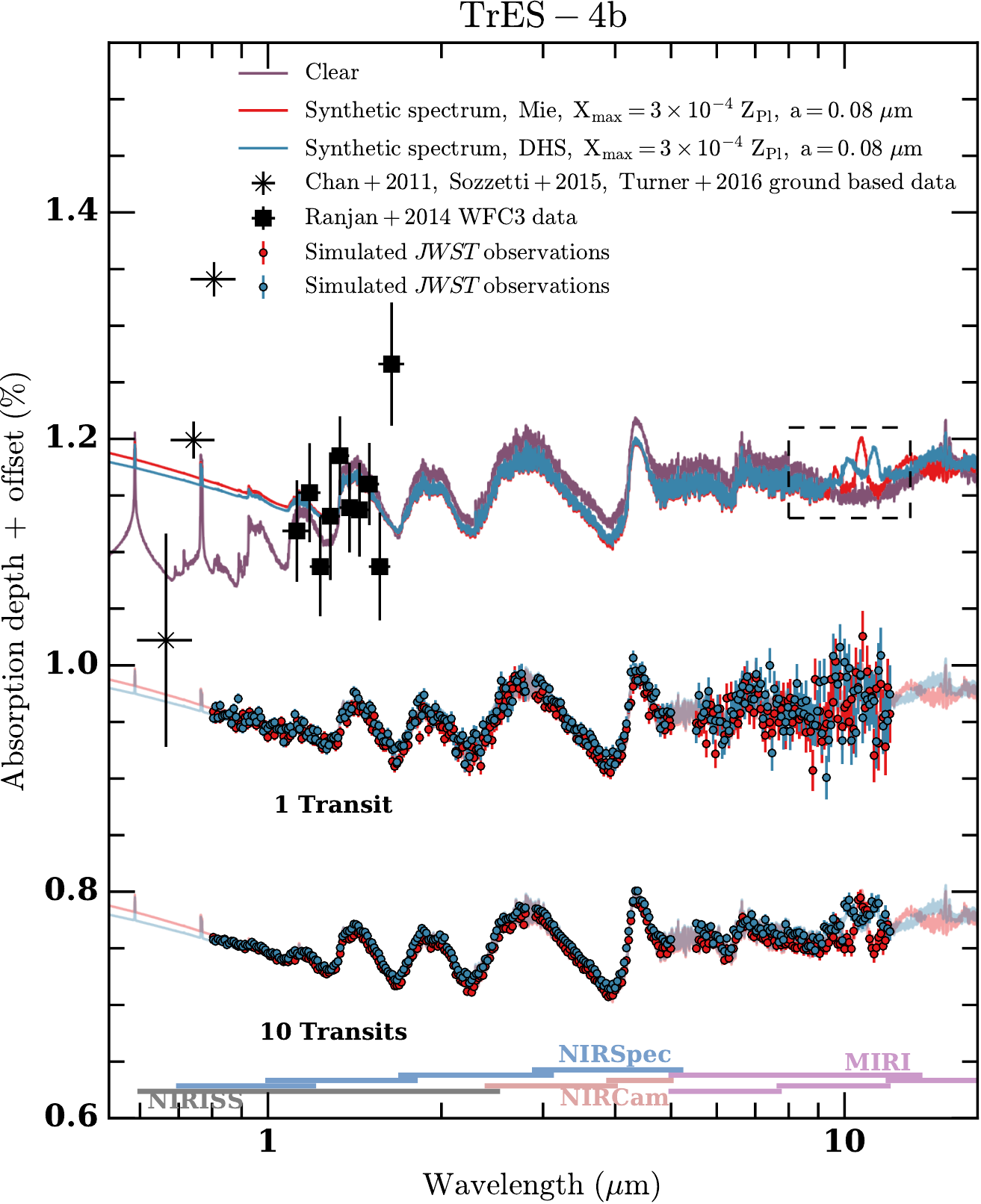}
\raisebox{0.01\height}{\includegraphics[width=0.486\textwidth]{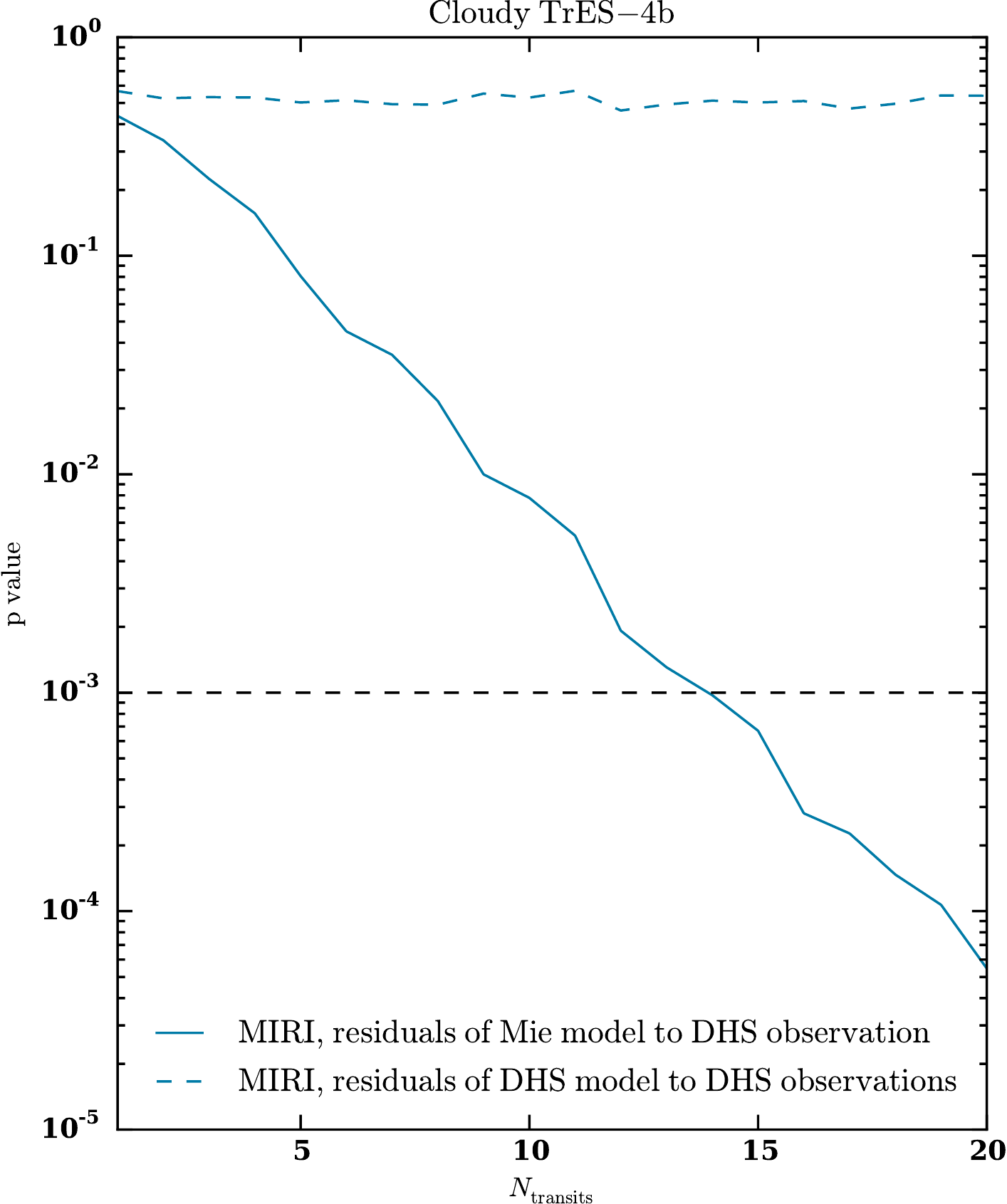}}
\caption{{\it Left panel:} synthetic transit spectra, observations, and synthetic observations for 
the planet TrES-4b. Black crosses denote the ground based observational data by 
\citet{chaningemyr2011,sozzettibonomo2015,turnerpearson2016}. Black squares denote the \emph{HST WFC3} data by \citet{ranjancharbonneau2014}. Synthetic spectra for for the homogeneously 
cloudy models with $X_{\rm max}=3\times 10^{-4}$~$Z_{\rm 
Pl}$ are shown as teal and red solid lines for the DHS and Mie opacity, 
respectively. The dashed box at $\sim$10~$\mu$m highlights the silicates features due to Mg$_2$SiO$_4$ resonances.The teal and red dots show the corresponding synthetic 
observations derived for 1 and 10 transits, re-binned to a resolution of 50. Vertical offsets have been applied for the sake of clarity. {\it Right panel:} $p$~values of the Kolmogorov-Smirnov test of the residuals between the synthetic observation of the TrES-4b DHS model and the Mie model (solid teal line). The $p$~value is shown as a function of $N_{\rm transit}$ for data taken with \emph{MIRI LRS} considering only the wavelength range of the silicate feature (9-13~$\mu$m) and correcting for global model offsets. The dashed teal line shows the $p$~value obtained when analyzing the residuals of the DHS model to its own observation. The black dashed line denotes our threshold value of $10^{-3}$.}
\label{fig:tres_4_transm}
\end{figure*}
 
Finally, we find that our KCl+Na$_2$S cloud model presented for GJ~1214b in the left panel of Figure \ref{fig:gj1214_transm} results in a $p$~value of 0.45 if compared to the existing observational data. It is therefore consistent with these data.

\subsubsection{TrES-4b}

In the left panel of Figure \ref{fig:tres_4_transm} we show the simulated 
transmission observations of \rch{TrES-4b}. \rch{TrES-4b is a strongly inflated (1.7 $\rj$) hot jupiter that circles its F-type host star ($T_*=6200~K$) once every 3.6 days \citep{chaningemyr2011}. The dayside emission of this planets seems to be consistent with a temperature inversion \citep{knutsoncharbonneau2009} and the dayside for this planet may therefore be too hot to have a significant silicate cloud coverage. However, the limbs may be much cooler than the bulk dayside, allowing for these clouds to exist \citep[see, e.g.,][]{Wakeford:2016ut}. To account for this effect, we model the transmission spectra for the planets assuming a global redistribution of the stellar irradiation energy (also see Section \ref{sect:clouds_spec}). The theoretical global equilibrium temperature of this planet is 1795~K. It is therefore within the temperature range where mineral clouds such as Mg$_2$SiO$_4$ and MgAl$_2$O$_4$ can be expected.}

We here look at synthetic observations for the cloud models 6 and 9 in Table \ref{tab:cloud_models}, i.e. homogeneously distributed clouds of small particles assuming either irregular (DHS) or spherically homogeneous (Mie) particles. For comparison we also show the clear model for this planet.

\rch{The feature at 10~$\mu$m in the cloudy models, which is highlighted by the dashed-line box in Figure \ref{fig:tres_4_transm}, arises from resonances of the crystalline Mg$_2$SiO$_4$ particles. The differences in the location and relative strength of the Mg$_2$SiO$_4$ resonance peaks, arising from the different particle shapes (irregular vs. spherically-homogeneous), are evident.}

\rch{By carrying out a Kolmogorov-Smirnov analysis of the residuals between the clear model and the synthetic DHS cloud model observations we find that a single transit in \emph{MIRI} will be enough to discriminate between the clear and cloudy model. This implies that a single transit is sufficient to find evidence for silicate cloud particles in such an atmosphere. Before carrying out the Kolmogorov-Smirnov analysis we made sure to correct the offset between the two models in the MIR part shortward of 9 $\mu$m, which is fully determined by molecular features. This was done in order to prevent a model discrimination based purely on global model offsets.}

While the existence of an extinction feature at 10~$\mu$m, spanning the 
wavelength range from $\sim$8 to $\sim$12~$\mu$m, would hint at 
the presence of silicate absorbers in the planet's atmosphere, a single transit will not be enough to discriminate between all possible silicate absorbers: The resolution needed to resolve individual crystalline dust features is in the range of 50. Thus the number of stacked transits needs to guarantee a high enough SNR for a single point at this resolution.
\citet{juhaszhenning2009} have shown that 
for protoplanetary disks the SNR required to characterize the silicate dust 
properties (crystallinity and size) well is between 10 and 100 per spectral 
point.
In the example shown here we only consider crystalline Mg$_2$SiO$_4$ for the silicates.
But also different silicates such as MgSiO$_3$, iron-enriched olivines and pyroxenes, or species such as SiO$_2$, FeSiO$_3$ and Fe$_2$SiO$_4$ are possible \citep[see][]{wakefordsing2015}. Another complication arises from the wavelength dependent shape and position of the 
absorption features for particles larger than $\sim 1$~$\mu$m, \rch{but note that 
a strong Rayleigh signal observed in the optical and NIR transmission 
spectrum would suggest particles which are smaller than 0.1~$\mu$m.
Unfortunately, the currently available ground based data for this planet exhibits a large spread such that conclusive statements regarding the optical and NIR part of the planet's spectrum appear difficult.}

\rch{In summary this} means that a single transit is not enough to fully characterize silicate dust based on the 10~$\mu$m feature in transmission spectra. However, if the need for small particle 
clouds is evident from the transmission spectrum, due to a strong Rayleigh 
signal in the optical and NIR, then the observation of a 10 $\mu$m feature 
presents strong evidence for the presence of silicate grains in the atmospheres, while the lack of 
such a feature means that the strong Rayleigh slope in the optical and NIR 
cannot be caused by silicates. In that sense \emph{JWST} will shed light on the nature of small grain clouds by allowing us to find, 
potentially using a single transit observation with \emph{MIRI}, whether 
silicates are responsible or not.

In our idealized example shown here, where the only considered silicate 
species are crystalline Mg$_2$SiO$_4$ particles of either irregular or spherically-homogeneous shape we can now asses how many transits would be needed to conclusively distinguish between both models. Similar to the Kolmogorov-Smirnov test carried out \rch{before for the clear and cloudy model} one can analyze the residuals of the Mie cloud model to the observations of the DHS cloud model. The results are shown in the right panel of Figure \ref{fig:tres_4_transm}. \rch{Again,} because the silicate features only occur in the \emph{MIRI} wavelength regime we concentrate only on this instrument for the analysis and only on the silicate feature, considering a wavelength range from 9-13~$\mu$m. In order to analyze only the difference in the cloud features we again corrected the cloud models for the offset which can be seen in the \emph{MIRI} wavelength regime, outside of the silicate feature.

We find that the Mie cloud model is inconsistent with DHS observations if \rch{$\sim$13-14} transits are stacked for this planet.
Because we only considered 2 possible cloud models here, and not the full parameter space, the number of transits needed to characterize the state of the clouds may likely be higher, however. \rch{Again, looking at the 10~transit observation in Figure \ref{fig:tres_4_transm} a discrimination between the 2 cases seems to be possible already at a smaller number or transits. Therefore, similar to the analysis carried out for GJ~1214b in Section \ref{subsubsect:gj1214b}, we expect that statistical tools more powerful than the ones used here should be able to retrieve this difference already for a smaller number of transits.}

The two cloudy models shown in Figure 6 are consistent with the observational \emph{HST} and \emph{Spitzer} data, resulting in $p$~values of \rch{0.20} an.d \rch{0.25} for the DHS and Mie cases, respectively.

\begin{figure*}
\centering
\includegraphics[width=0.485\textwidth]{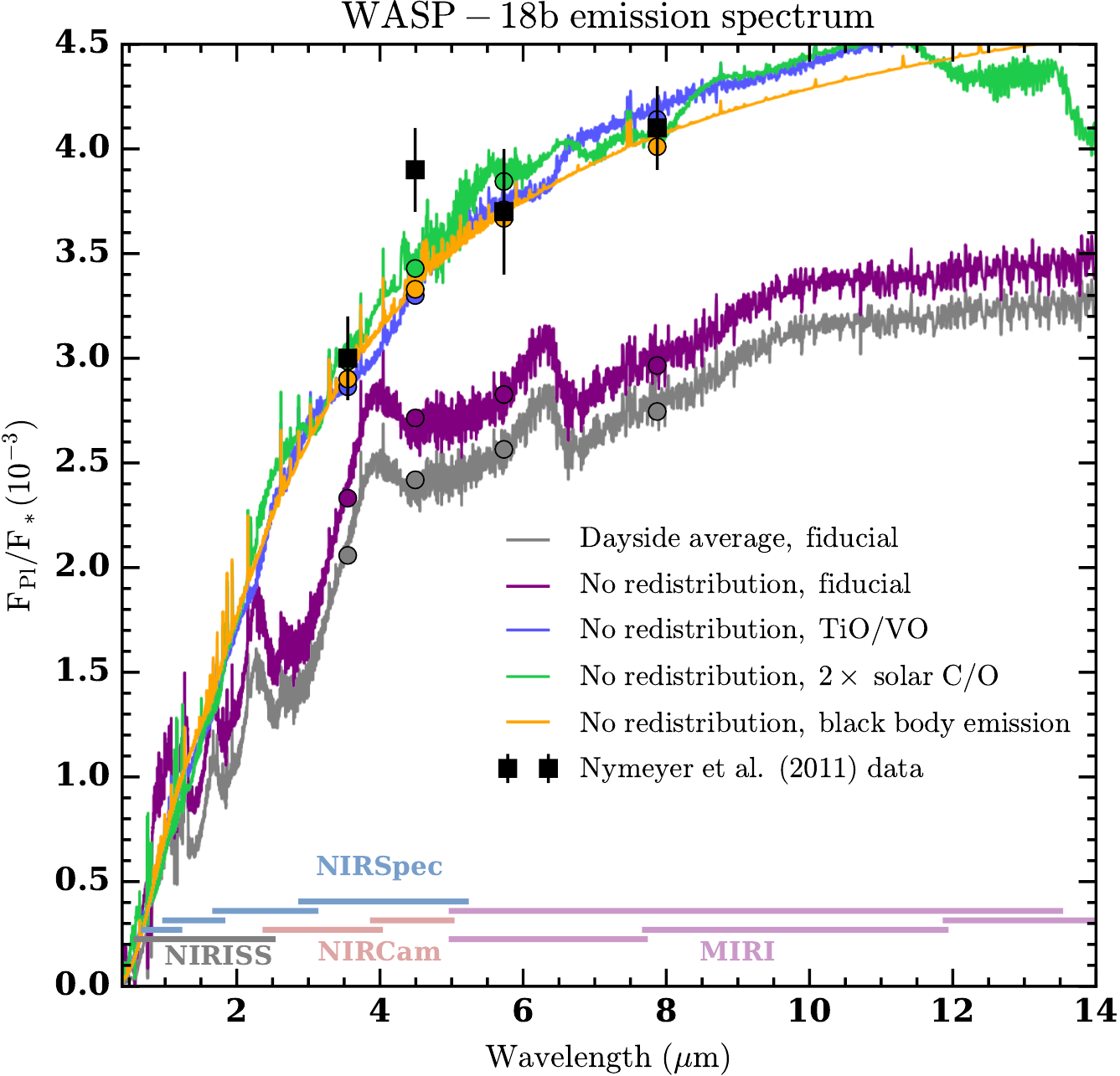}
\raisebox{0.01\height}{\includegraphics[width=0.465\textwidth]{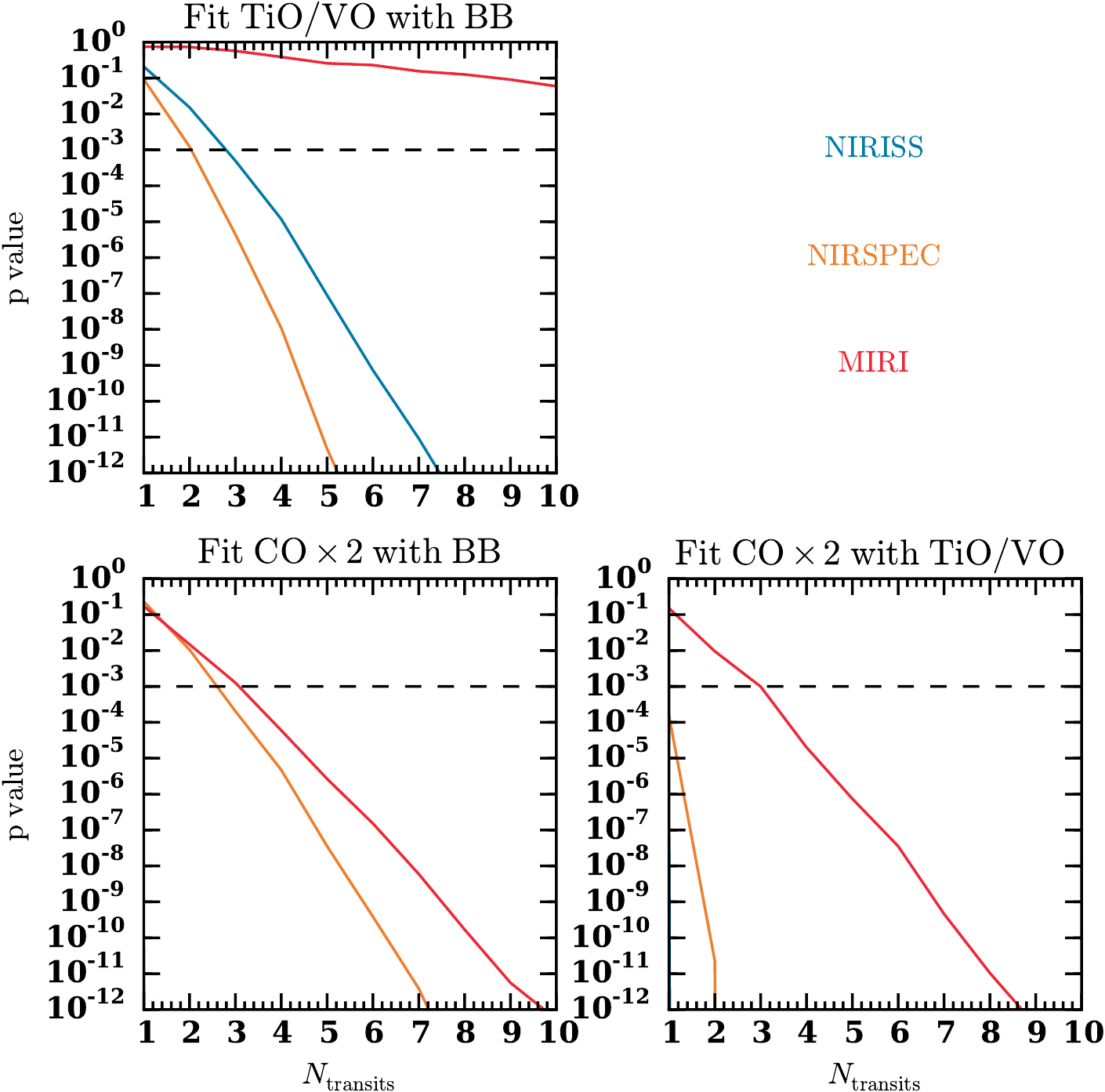}}
\caption{{\it Left panel:} Synthetic spectra and observational data of the emission spectrum of WASP-18b. A description of the different lines is shown in the legend. The observational data by \citet{nymeyerharrington2011} are shown as black errorbars. The ``emission'' lines which can be made out in the blackbody $F_{\rm Pl}/F_{\rm *}$ spectrum are absorption lines of the stellar spectrum. \rch{The colored circles show the corresponding \emph{Spitzer} channels for the synthetic spectra.} {\it Right panel:}
$p$~values of the Kolmogorov-Smirnov test of the residuals between the synthetic observation of the 3 WASP-18b models and the 2 respective remaining models. The $p$~value is shown as a function of $N_{\rm transit}$ for data taken with the 3~different instruments listed in Table~\ref{tab:JWST_parameters}. In order to avoid distinguishing the models based on offsets the residual distributions were shifted to have a mean values of 0. The black dashed line denotes our threshold value of $10^{-3}$. If an instrument is not shown in one of the 3 subpanels then it has a $p$~value lower than 10${-12}$ already after 1 observation.}
\label{fig:wasp_18_no_redist}
\end{figure*}

\subsection{Emission spectroscopy}
In this section we will investigate simulated eclipse observations of our 
targets, using \emph{JWST}. We concentrate on a very interesting class of 
emission targets, namely the hottest hot Jupiters in our target selection: 
WASP-33b, Kepler-13Ab and WASP-18b. Along with their high equilibrium 
temperatures, these planets share an additional similarity: Their spectra can 
all be approximated well by blackbody emission, yet all of them are best fit by 
inversions in their atmospheres, see \citet{haynesmandell2015}, 
\citet{shporerorourke2014} and \citet{nymeyerharrington2011} for WASP-33b, 
Kepler-13Ab, and WASP-18b respectively. Further examples for such planets 
in the literature are TrES-3b \citep{crolljayawardhana2010} and WASP-24b 
\citep{smithanderson2012b}.

Especially the case of WASP-18b is intriguing: Based on the orbital and stellar 
parameters, the planet's theoretical equilibrium temperature (i.e. assuming a planet-wide average of the insolation) is $T_{\rm equ}
=2410$~K. If one assumes a dayside averaging of the insolation flux then the 
dayside effective temperature $T_{\rm irr}$ would be 2870~K. Yet, the 
measured dayside emission flux of this planet is consistent with brightness 
temperatures between 3100 and 3300~K, and the spectrum can be fit 
reasonably well by a blackbody at 3200~K \citep[although an inversion fits 
better, see][]{nymeyerharrington2011}.
Theoretically, the maximum flux which can be measured for an irradiated 
planet when observed during transit geometry, and when assuming blackbody 
emission, is equal to the flux emitted by a blackbody of temperature $T_{\rm 
max}=T_*\sqrt{R_*/a}$, where $T_*$ is the stellar effective temperature, $R_*
$ the stellar radius and $a$ the planet's semi-major axis. Note that the shape 
of the SED of such a planet would correspond to an even higher temperature, 
because the planet must emit all flux close to the substellar point and into the 
direction of the observer: if the planet truly had a global temperature of 
$T_{\rm max}=T_*\sqrt{R_*/a}$ it would violate energy conservation, emitting 
more flux than it receives from the star.

For WASP-18b $T_{\rm max}=3410$~K, such that the planet's corresponding 
blackbody temperature of 3200~K is still below this theoretical limit. 
Nonetheless it suggests that a non-negligible fraction of the planetary flux 
must be emitted close to the location of the substellar point and thus into the 
direction of the observer during eclipse geometry. One can also see this by considering the insolation flux received 
by each circular planetary annulus at an angle $\theta$ away from the 
substellar point, which is $F(\theta)=T_*^4(R_*/a)^2{\rm cos}(\theta)$. If one neglects any 
energy redistribution due to winds then the planetary surface at angle $\theta$ away from the substellar point has to reemit exactly $F(\theta)$. Assuming blackbody emission one finds that the flux measured during transit geometry is the same as if the 
planet had a global temperature of $T_{\rm rad}=(2/3)^{1/4}T_*\sqrt{R_*/a}$, 
which corresponds to 3080~K for WASP-18b. This is less than the stated 
blackbody temperature of 3200~K, suggesting that for this planet energy 
redistribution may not only be limited, but fully absent.

The same situation seems to be the case for Kepler-13Ab, which is best fit by 
an inversion in its atmosphere, yet also reasonably well described by a 
blackbody at 2750~K \citep{shporerorourke2014}. For this planet the 
theoretical upper limit on the observable effective temperature as seen during 
eclipse geometry is $T_{\rm max}$=3085~K, the temperature derived for the 
case fully neglecting energy redistribution is $T_{\rm rad}$=2790~K, the 
dayside averaged effective temperature is $T_{\rm irr}$=2590~K if nightside 
emission is neglected and the global equilibrium temperature would be 
$T_{\rm equ}$=2180~K. The measured effective temperature of the planet is 
closest to $T_{\rm rad}$, suggesting that wind redistribution of stellar 
insolation energy may be neglected for this planet as well. Interestingly, and 
quite paradoxically, \citet{shporerorourke2014} derive a geometric albedo of 
$0.33^{+0.44}_{-0.06}$ for this planet, corresponding to a bond albedo of 0.5, 
if a matte, i.e. perfectly Lambertian scattering process is assumed. This 
albedo value arises from the fact that the optical eclipse depth for this planet 
shows an excess which cannot be explained by 1d model calculations 
investigated in \citet{shporerorourke2014}. The effective blackbody 
temperature derived from emission observations in the infrared (2750~K) is inconsistent with a bond 
albedo of 0.5, for which $T_{\rm max}$ would be only 2590~K. Note, however, 
that the derivation of the geometric albedo in \citet{shporerorourke2014} 
assumed that the brightness temperature as well as the geometric albedo are 
constant within the 3 different bands used in their analysis, which is not 
necessarily the case. Further, the presence of scattering aerosols in the 
planet's atmosphere in this high albedo case requires particles which are 
stable even at the high temperatures found for this planet, which is 
challenging. The measured excess of the optical eclipse depth may therefore be the emission feature of an unknown opacity source.

The planet WASP-33b is less extreme, because its theoretical dayside averaged effective 
temperature is $T_{\rm irr}$~=~3250~K, i.e. still above the value derived when a blackbody is 
fitted to the emission observations \citep[2950~K, see][]{haynesmandell2015}. 
Note that also this planet is fit better by an inversion than by a blackbody.

Given the fact that WASP-18b and Kepler-13Ab seem to have 
only a limited, or no redistribution of the stellar insolation energy at all we decided to calculate spectra for these planets fully neglecting the redistribution using Scenario (iii) described in Section \ref{sect:irrad_treatment}.

We show the resulting spectra in the left panel of Figure 
\ref{fig:wasp_18_no_redist}, together with the data by 
\citet{nymeyerharrington2011}. No offset or scaling factor was applied to the synthetic spectra. As one can see, our fiducial case, i.e. the case 
without clouds, solar C/O ratio and no TiO/VO opacities in its atmosphere does the 
worst job at fitting the data, being multiple sigmas away from all the measured 
points. For comparison we also plot the fiducial case when assuming a dayside averaged insolation. In this case the synthetic spectrum is even further away from the data.

Only three scenarios provide a good fit to the data, namely the case where we included TiO/VO opacities, which lead to an inversion, the case where we consider a C/O which is twice the solar value, leading to C/O~=~1.12, and the case where we assume the planetary annuli to emit as isothermal blackbodies. It was crucial to neglect redistribution for these cases. The corresponding dayside averaged cases resulted in fluxes which were too low. The ``emission'' lines which can be made out in the blackbody $F_{\rm Pl}/F_{\rm *}$ spectrum are absorption lines of the stellar spectrum.

\rch{Note that the \emph{Spitzer} point at 4.5~$\mu$m does not seem to be fitted by the models. However, a good fit to data is not about the model perfectly describing every data point. In fact, if the error bars of a measurement are estimated accurately, then one expects that 1/3 of all measured points are further than 1~$\sigma$ away from the prediction of the ``correct'' model. To assess the goodness of fit of the models to the data we will again make use of the Kolmogorov Smirnov test. Note that $\chi^2$ may be used to compare the various models against each other, but not to assess the overall goodness of fit of a given model: For a linear model, the expected value of a model correctly describing the data is $\chi^2=\#{\rm dgf}$, where $\#{\rm dgf}$ is the number of the degrees of freedom, such that $\chi_{\rm red}^2=1$. However, the spectral models we use here are non-linear, with the exact number of the degrees of freedom unknown. The expected $\chi^2$ value of a model consistent with the data can therefore not be calculated, and the use of the $\chi^2$ to assess the goodness of fit is not allowed \citep[also see][]{andrae2010}. The $p$~value of the Kolmogorov Smirnov test applied on the residuals between the WASP-18b \emph{Spitzer} measurements and the data, on the other hand, represents a valid method for assessing the goodness of fit. The $p$~values for the TiO/VO, C/O=2$\times$(C/O)$_\odot$ and blackbody model are 0.38, 0.82 and 0.13, respectively. All models are therefore consistent with the data, with the best fit being provided by the C/O=2$\times$(C/O)$_\odot$ model.}

For the high temperatures considered here gaseous SiO may become important because the dayside of the planet is too hot to form any silicate clouds. We neglect the opacity of SiO, but this molecule is a strong UV absorber for $\lambda<0.3$~$\mu$m \citep[see, e.g.,][]{sharpburrows2007}  and may therefore lead to even stronger inversions. In future calculations an inclusion of the SiO opacities for these hottest planets is therefore necessary.

The C/O~=~1.12 case fits the data well because C/O ratios close to 1 may cause inversions, see \citet{mollierevanboekel2015}. Such inversions form because for C/O ratios close to 1 oxygen and carbon are predominantly locked up in CO, decreasing the abundance of water if approached from C/O~$<$~1, or that of methane if approached from C/O~$>$~1. Because water and methane have large IR opacities, they are the atmosphere's most effective coolant. Therefore, for C/O~$\sim$~1 the cooling ability of the atmospheres is decreased, while the heating due to the alkali absorption of stellar light stays  strong (note that we include equilibrium ionization for sodium and potassium).

Given the fact that the blackbody emission case fits the data well one can also understand why the inversion cases provide a good fit to the data: the inversion stops the monotonous decrease of temperature within the atmosphere, leading to smaller temperature variations across the photosphere. The photosphere therefore becomes more isothermal \rch{(also see Section \ref{sect:tiovo_behavior})}.

Similar to the analyses carried out for the transiting planets we will now look into the number of transits needed to distinguish between the three models which fit the observational data best. Applying an offset to the spectra for the emission spectra presented here would violate energy redistribution such that we will carry out the analysis without applying an offset at first. In this case the spectra of the 3 models can be distinguished from each other using only a single eclipse observation, in either of the three instruments considered here. 

However, in order to assess how well the models may be distinguished because of differences in their spectral shape we next applied an offset to the models before carrying out the Kolmogorov-Smirnov test. This was done by shifting the mean value of the residual distribution to zero. The corresponding plots are shown in the right panel of Figure \ref{fig:wasp_18_no_redist}. Physically, such an offset may be motivated by a non-negligible redistribution of the stellar insolation, decreasing the planetary flux measured during an eclipse observation.
Note that a simple offset as carried out here is only an approximation because the spectral shape may change under such conditions.

In general the results follow our expectations: \emph{JWST} is less capable to distinguish between the TiO/VO and the blackbody case because the TiO/VO case does not have many features and is mostly simply offset in comparison to the blackbody model. This is especially true in the \emph{MIRI} wavelength range, such that 10 transits are not enough to distinguish between the two models at high confidence, because we do not allow for the distinction between two models just because of an offset: this is why we applied the aforementioned shift of the residual distribution. The instrument best suited for distinguishing both models is \emph{NIRSpec}, achieving this goal in just 4 transits.

\emph{JWST}'s capability to distinguish the CO~=~1.12 and the blackbody case is much better, due to features visible in the spectrum of the CO=1.12 case:
The features at 3~$\mu$m, 6.5 to 8.5 and from 11~$\mu$m onward all stem from HCN {\it absorption} (not emission), whereas the feature at 4.5 $\mu$m is caused by CO absorption. These are typical absorbers expected for a hot, carbon-rich atmosphere. Because the \emph{NIRISS} wavelength range is devoid of any molecular features observations in here will not help to distinguish between the models. The largest diagnostic power is provided by using \emph{NIRSpec} observations (1 eclipse measurement) while \emph{MIRI} can discriminate between the models after 3 eclipse measurements.

The easiest case to distinguish is the case when comparing the \rch{C/O} = 1.12 to the TiO/VO model because both these cases show molecular features. Again \emph{NIRSpec} is best at achieving this goal, using just a single transit, whereas \emph{MIRI} and \emph{NIRISS} need 2 and 3 transits, respectively.
In conclusion one can therefore say that if one of our self-consistent models was the true state of the atmosphere, then \emph{JWST} could determine its state by taking 1-4 transits in \emph{NIRSpec}. Of course we do not prove that the three models here are the only possible ones, but our example illustrates the foreseen diagnostic power of \emph{JWST }for such atmospheres.

\section{Format and extent of the published atmospheric structures, synthetic spectra and observations.}
\label{sect:data_pub}
For all target planets listed in Table \ref{tab:obs_cands} we publish the self-consistent atmospheric structures listing the pressure, temperature, density, mean molecular weight and specific heat $c_P$ for each atmospheric layer. We also publish the mass and number fractions of all chemical species as well as the mass fractions of all cloud species derived from the cloud models and the mean cloud particle sizes for every layer.

Further, we publish emission model spectra, reflected light spectra and transmission spectra from 110~nm to 250~$\mu$m. Additionally we publish spectra restricted and rebinned to the \emph{JWST} wavelength range and intrinsic resolution of the instruments listed in Table \ref{tab:JWST_parameters}. For these rebinned spectra we publish single transit/eclipse errorbars, such that observations at any number of transits may be obtained by Gauss-sampling the noiseless spectra using errorbars scaled with $1/\sqrt{N_{\rm transit}}$. Lower resolution observations can be obtained by rebinning the observations and propagating the observational errors.

These data are published for all scenarios considered for the target planets, i.e. varying the enrichment, C/O ratio, irradiation treatment, cloud model, including TiO/VO opacities etc. 

We publish all these data as described above in order to enable users to reproduce our calculations if desired and to be able to study the calculations as well as possible. Tests regarding the observational distinguishability via the Kolmogorov-Smirnov method as introduced in Section \ref{sect:sim_obs} are possible with the data as well as testing whether or not retrieval codes are able to constrain the atmospheres parameters as given in our published data.

\section{Summary and conclusion}
\label{sect:summ_and_conc}
In this study we present a set of self-consistent atmospheric calculations for prime transiting exoplanet targets to be observed with \emph{JWST}. We publish the resulting atmospheric structures, abundances and transmission and emission spectra. For the spectra we additionally publish wavelength dependent uncertainties for \emph{JWST} observations, derived from radiometric modeling. By sampling the noiseless data using these errors synthetic observations can be obtained.

The exoplanet targets have been chosen because they have a high expected signal-to-noise, cover the $T_{\rm equ}$--${\rm log}(g)$ space homogeneously, and include planet types ranging from super-Earths to hot Jupiters. This diverse set of targets may therefore allow to study the full breadth of transiting exoplanets at high SNR.

Because the data currently available for these planets is often limited both in its spectral coverage and SNR it is crucial to explore different atmospheric scenarios in order to assess the width of possible observational results which may be seen once \emph{JWST} becomes available. To this end we explore a wide range of scenarios by varying the planets' enrichment, composition (C/O ratio), optionally include absorbers in the optical (TiO/VO) and put a large emphasis on studying the effect of different cloud properties. Furthermore we apply different assumptions for the heat-redistribution; this we mimic by changing the irradiation, assuming either a planet-wide or dayside average of the irradiation. For some selected, very hot planets we also study the case of fully neglecting the redistribution of stellar irradiation.

Given the large uncertainties when trying to model clouds self-consistently we study two cloud model setups, using either our implementation of the \citep{ackermanmarley2001} cloud model, for which we publish a new derivation, or by applying a parametrized cloud model. In the latter model we set the cloud mass fractions equal to the condensate mass fraction obtained from equilibrium chemistry, but impose an upper boundary as a free parameter. Another free parameter is the particle size for the mono-disperse size distribution.
This model therefore represents the case of vertically homogeneous clouds with mono-disperse particle distributions.
Because cloud particles in planetary atmospheres are expected to be crystalline, rather than amorphous, we use optical constants for crystalline material whenever available. Additionally, all direct detections of crystalline silicate grains in different astrophysical contexts suggest that the grains should be irregularly shaped dust agglomerates, such that this is our standard assumption. The case of spherically-homogeneous cloud particles, with opacities derived from standard Mie theory, is studied as an optional case.

All calculations presented here are compared to observational data whenever available. For a selected subset of our targets we study and compare different models which are consistent with the available data and investigate how well \emph{JWST} may be at distinguishing these models, and which instrument is best used for this task.
The method we use for this subset study may be applied to any models within our grid. We summarize the main findings of this study below.
\begin{itemize}
\item {\bf Super-Earths with a thick cloud cover} were studied by investigating our models for GJ~1214b. We find that if we assume a heavy enrichment (1000~$\times$ solar) and a thick cloud deck then our models are consistent with current observational data. While the current data is consistent with a \rch{completely flat, featureless} spectrum we find that $<$10 transits with the \emph{NIRSpec} instrument may be sufficient to unambiguously reveal CO$_2$ and CH$_4$ features in the atmosphere of this planet. For a more conclusive statement on the number of transits needed to characterize this planet, given our model, more sophisticated statistical tools, such as retrieval analyses, are required. We plan to carry out such analyses as a next step.
\item {\bf Gas giants} were studied by investigating our models calculated for \rch{HAT-P-12b and TrES-4b}. In concordance with previous studies we find that vertically homogeneous, small particle ($<0.1$~$\mu$m) clouds \rch{are best at producing strong Rayleigh scattering signatures}, but only if iron-bearing cloud species are neglected. \rch{For TrES-4b we expect a feature at 10~$\mu$m in the transmission spectrum if it harbors clouds that are made up from such small silicate particles. We find that 1 transit with \emph{MIRI} may be sufficient to reveal the 10~$\mu$m feature, while less than 10 transits may be enough to distinguish between irregularly shaped or spherically-homogeneous cloud particles, if the silicate species is known. Similarly, more sophisticated statistical tools will improve the analysis of the minimum number of transits required.} A full characterization of silicate cloud particles \rch{(species, size distribution, vertical extent, particle shape, etc.)} will likely require more transits.
\item {\bf Extremely hot transiting planets} are often well fit by isothermal emission when comparing to the data from eclipse measurements. We study the planet WASP-18b as an example and find that self-consistent models can explain the observations of this planet if energy redistribution by winds is fully neglected. The model setups which fit current observational data best are models featuring inversions either because of TiO/VO absorption or because of C/O number ratios close to 1. In this latter scenario the main coolants (water or methane and HCN) are significantly depleted in favor of CO, such that inversion form. We find that a single eclipse observation with \emph{NIRSpec} is enough to distinguish these cases.
\end{itemize}

By investigating these three example cases we show that \emph{JWST} will be able to shed light on many intriguing puzzles of atmospheric studies which are difficult to solve using today's observational facilities. Further, by publishing our atmospheric model calculations along with synthetic observational uncertainties for \emph{JWST} we allow for the study of different possible scenarios and how well they can be observed and distinguished.

It has to be kept in mind that ruling out given models against each other does not answer how conclusively we will be able to characterize a given atmosphere using \emph{JWST} data. For such assertions retrieval studies for the synthetic models would have to be carried out, and the results compared to the input model, as was done in \citet{greeneline2016}, but even then the conclusions depend on the input models. Nonetheless, studying the atmospheric models for the target planets as presented here enables to vet the power of \emph{JWST} at constraining the atmospheric state given various likely, self-consistent solutions for the investigated planets, which are consistent with the data available today.

Finally, because the full models are published, including temperature and abundance structures, retrieval models may be tested on our grid, allowing to study the retrievability of the expected \emph{JWST} observations when considering self-consistent atmospheric models.

\begin{acknowledgements}
P.M. thanks J.-L. Baudino, B. Bezard, C. Dullemond, E. Ford, H.-P. Gail, D. Homeier, C. J\"ager, H.-G. Ludwig, C. Mordasini, V. Parmentier, P. Tremblin, and A. Wolfgang for useful discussions and 
helpful comments. POL acknowledges support from the LabEx P2IO, the French ANR contract 05-BLAN-NT09-573739.
\end{acknowledgements}
 
\appendix
\section{Updates of \petit}
\label{app:changes}
\subsection{Line cutoff} Previously the opacities used in \petit did not consider 
a sub-lorentzian cutoff of the line wings sufficiently far away from the line 
center. To include the effect of a line cutoff we use measurements 
by \citet{hartmannboulet2002} for all molecules but CO$_2$. The cutoff is 
modeled by means of an exponential line wing decrease. For CO$_2$ we 
make use of a fit to the CO$_2$ measurements by \citet{burchgryvnak1969}, 
which was obtained from Bruno Bezard (private communication). In 
\citet{hartmannboulet2002} CH$_4$ lines broadened by H$_2$ have been 
measured. Because measurements for other species different from CH$_4$ 
and CO$_2$ do not exist we use the CH$_4$ cutoff for all remaining species 
as well.

\subsection{Chemistry}
The chemical equilibrium abundances in the \emph{petitCODE} are now 
calculated with a self-written code that minimizes the Gibbs free energy, which was 
implemented closely following the methods and equations outlined in the 
\emph{CEA} manual \citep{gordon1994}. The code converges reliably 
between 60~-~20000~K. Moreover, it was checked for consistency with the 
\emph{CEA} code \citep{gordon1994,mcbride1996}, leading to excellent 
agreement in the temperature range for which the \emph{CEA} 
thermodynamic data are valid.

For condensed material with no available thermodynamic data at cold 
temperatures the heat capacity $c_P$ was extrapolated to low temperatures 
by fitting a Debye curve to the higher temperature (usually $>$~300~K) $c_P$ 
data, assuming $c_V = c_P$ for the solid material:
\beq
c_P \propto \left(\frac{T}{T_D}\right)^3\int_{0}^{T_D/T}\frac{x^4e^x}{(e^x-1)^2}
dx \ ,
\eeq
where $T_D$ is the Debye temperature. The fitted function could then be used 
to obtain $c_P$ at low temperatures.
The entropy $S$ and enthalpy $H$ were obtained using $dS = c_P T^{-1} dT$ 
and $dH = c_P dT$.
The thermodynamic data used for the solids were either the ones used in the 
\emph{CEA} code\footnote{\url{http://www.grc.nasa.gov/WWW/CEAWeb/
ceaThermoBuild.htm}}, the data given in the \emph{JANAF} database 
\footnote{\url{http://kinetics.nist.gov/janaf/}} or data described in 
\citet{robie1978}.

The condensible species which the code can currently treat are Al$_2$O$_3$, Fe, Fe(l), FeO, Fe$_2$O$_3$, Fe$_2$SiO$_4$, H$_2$O, H$_2$O(l), H$_3$PO$_4$, H$_3$PO$_4$(l), KCl, MgSiO$_3$, MgSiO$_3$(l), Mg$_2$SiO$_4$, Mg$_2$SiO$_4$(l), MgAl$_2$O$_4$, Na$_2$S, SiC, SiC(l), TiO, TiO(l), TiO$_2$, TiO$_2$(l), VO and VO(l), where the phase of all 
species is solid unless its name is followed by an ``(l)'', which stands for liquid 
phase.

\subsection{Clouds}
\label{sect:app_cloud_mod}
For the cloud module we implemented the model as introduced and described 
by \citet{ackermanmarley2001}, for which one needs to solve the equation
\beq
K\frac{\partial X_{\rm t}}{\partial z} + f_{\rm sed}v_{\rm mix}X_{\rm c} = 0 \ ,
\label{equ:ack_marley_main}
\eeq
where $K$ is the local atmospheric eddy diffusion coefficient, $X_{\rm t}$ is 
the total mass fraction (condensate + gas) of the cloud species, $v_{\rm mix}$ 
is the local atmospheric mixing velocity (arising from diffusion), $X_{\rm c}$ is 
the condensate mass fraction of the cloud species and
\beq
f_{\rm sed} = \frac{\left<v_f\right>}{v_{\rm mix}} \ ,
\eeq
with $\left<v_f\right>$ being the mass averaged settling velocity of the cloud 
particles in a given layer.

{\it We found that Equation \ref{equ:ack_marley_main} is correct independent 
of any cloud nucleation, condensation, coagulation and shattering processes, 
as long as the transport mechanism of cloud particles and the gas is diffusive.} 
We attach a derivation of this statement in Appendix \ref{app:cloud_deriv}. 
Limiting assumption of this model are that the particle distribution within the 
cloud follows a log-normal distribution with width $\sigma = 2$, that clouds of 
different species cannot interact, the assumption that all clouds within an 
atmosphere can be described using a single value of $f_{\rm sed}$ and that 
the cloud is forming in a diffusive environment, as opposed to a pure updraft 
as is assumed in, e.g., \citet{zsomkaltenegger2012}.

Note that we only allowed for the formation of a single cloud layer per species, 
which is then assumed to effectively act as a cold trap, preventing the 
formation of second, higher altitude cloud layers.

For solving Equation \ref{equ:ack_marley_main} within the framework of our 
code we rewrite is as
\beq
K\frac{\partial X_{\rm c}}{\partial z}+ f_{\rm sed}v_{\rm mix}X_{\rm c} = - K
\frac{\partial X_{\rm c}^{\rm equ}}{\partial z} \ ,
\label{equ:cloud_struct_final}
\eeq
where $\partial X^{\rm equ}_{\rm c}/\partial z$ is the gradient of the equilibrium 
chemistry condensate mass fraction.
A derivation of this form of the equation can be found in Appendix 
\ref{app:cloud_deriv}.

For obtaining the vertical eddy diffusion coefficient we assumed, analogous to 
\citet{ackermanmarley2001}, that there is a minimum $K_{\rm min} = 
10^5$~cm$^2$~s$^{-1}$ stemming from the breaking of gravitational waves in 
the atmosphere. Additionally we included two more contributions:

In the deep atmospheric layers, just above the convectively unstable region, 
we account for the motion arising from convective overshoot. To arrive at a 
simple description for the overshoot eddy diffusion coefficient we considered 
the fit reported in \citet{ludwigallard2002,hellingackerman2008}, namely
\beq
K_{\rm overshoot}(P) = K_{\rm MLT}\left[\frac{H(P)}{H_{\rm MLT}}
\right]^2\left(\frac{P}{P_{\rm MLT}}\right)^{\alpha g_5^{1/2}}\ ,
\eeq
where $K_{\rm MLT}$ is the eddy diffusion coefficient found in the last deep 
convective layer, i.e. just before the atmosphere becomes stable against 
convection further above. We set the mixing length $L=H$, where $H$ is the 
pressure scale height. $P_{\rm MLT}$ is the pressure in the last (uppermost) 
convectively unstable atmospheric layer and $P<P_{\rm MLT}$. The exponent 
terms are defined as $g_5 = g / (10^5\ {\rm cm}\ {\rm s}^{-2})$, where $g$ is 
the gravitational acceleration in the atmosphere. The $\alpha$ factor is a linear 
function of the internal temperature and varies between 1 und 3 for $T_{\rm 
int}$ ranging from 1500 to 300 K, following \citet{hellingackerman2008}.
To obtain $K_{\rm MLT}$ we implemented mixing length theory (MLT) as 
described in, e.g., \citet{kippenhahnweigert1990} and used
\beq
K_{\rm MLT} = \frac{H}{3}\left(\frac{RF_{\rm MLT}}{\mu\rho c_P}\right)^{1/3} \ ,
\eeq
with $R$ being the universal gas constant, $\mu$ the molecular weight in 
units of g per mol, $c_P$ the specific heat and $F_{\rm MLT}$ the energy flux 
in the atmosphere transported by convection.

We found that for many planets a self-consistent coupling of the overshoot 
mixing and the atmospheric temperature iteration lead to non-convergence. 
The reason is that in cases where clouds form deep in the atmosphere just 
above the convective region they can make the atmosphere sufficiently 
optically thick to trigger convection. This moves $P_{\rm MLT}$ to lower 
values. In the regions which then switch to being convective the increased 
mixing strength results in larger cloud particle radii (for a fixed $f_{\rm sed}$), 
which lead to a smaller cloud opacity which causes the culprit layers to 
become stable against convection again. These layers therefore oscillate 
between being convectively stable or unstable, impeding convergence. To 
circumvent this problem we decided to impose the overshoot mixing 
coefficient and thus set
\beq
K_{\rm overshoot}(P) = 10^9 \cdot \left(\frac{P}{1000{\rm \ bar}}\right) \ 
{\rm cm}^2 \ {\rm s}^{-1}\ ,
\eeq
which we found to be broadly consistent with the self-consistent values 
obtained for the various planets we considered in this work.
Note that this treatment is only valid for irradiated planets with atmospheric 
structures dominated by insolation as for self-luminous planets the radiative 
convective boundary moves to smaller pressures.

In the upper regions of irradiated planets one finds an increase of the eddy 
diffusion coefficient as the insolation drives vertical motion in the atmosphere 
and less stellar flux has been absorbed in the upper regions of the planet.
\citet{parmentiershowman2013} found in GCM models that the corresponding 
eddy diffusion coefficient behaves roughly as
$K_{\rm irrad} \propto P^{-1/2}$ and two GCMs modeling HD 189733b and HD 
209458b have found
\begin{align}
K_{\rm 209458b} & = 5 \times 10^8 \cdot \left(\frac{P}{1{\rm \ bar}}\right)^{-0.5} 
{\rm cm^2 \ s}^{-1} \ , \\
K_{\rm 189733b} & = 10^7 \cdot \left(\frac{P}{1{\rm \ bar}}\right)^{-0.65} {\rm 
cm^2 \ s}^{-1} \ ,
\end{align}
see \citet{agundezparmentier2014}.
We adopted an irradiation contribution to $K$ proportional to $P^{-1/2}$, 
where the reference value at 1~bar for HD 189733b was used. The difference 
between the mixing strength of HD 189733b and HD 209458b originated in 
the inclusion of TiO/VO opacities for HD 209458b, and similar values were 
obtained for HD 209458b if these opacities were neglected (private 
communication with V. Parmentier). For pressures smaller than $10^{-5}$~bar 
we held the $K_{\rm irrad}$ value constant to the value at $10^{-5}$~bar, 
because this is where the GCM calculation stops.

The full eddy diffusion coefficient is thus found as
\beq
K = {\rm max}\left(K_{\rm min}, K_{\rm overshoot} + K_{\rm irrad}\right) \ .
\eeq 
The mixing velocity can be obtained from $v_{\rm mix} = K/H$.

\subsection{Cloud opacities}
\label{sect:ref_cl_opa}
We calculate cloud opacities for (i) homogeneous spheres and for (ii) 
irregularly shaped cloud particles. Applying two different cloud particle 
treatments may allow for the distinction between spherical and irregular cloud 
particles in the case of small enough grain sizes for which the cloud 
material's resonance features are most clearly visible 
\citep{minhovenier2005}.
We approximated the opacity of the irregularly shaped cloud particles by 
taking the opacities obtained for a distribution of hollow spheres (DHS). The 
cross-sections for the spherical and DHS cloud particles were calculated using 
the dust opacity code of \citet{minhovenier2005}, which makes use of software 
reported in \citet{toonackerman1981}. The code uses Mie theory for the 
homogeneous spheres and an extended Mie formulation to take into account 
the hollowness of grains for DHS. As of now we include MgAl$_2$O$_4$, 
MgSiO$_3$, Mg$_2$SiO$_4$, Fe, KCl and Na$_2$S clouds with the real and 
complex parts of the refractive indices taken from \citet{palik2012} for MgAl$_2$O$_4$, \citet{scottduley1996,jaegermolster1998} for MgSiO$_3$, 
\citet{servoinpiriou1973} for Mg$_2$SiO$_4$, \citet{henningstognienko1996} 
for Fe, \citet{palik2012} for KCl and \citet{morleyfortney2012} for Na$_2$S. 
For the particles which have their optical properties described by DHS we use 
a porosity of $P=0.25$ \citep[as in][]{woitkemin2016} and an irregularity 
parameter $f_{\rm max}=0.8$ as defined in \citet{minhovenier2005}.

\subsection{TiO \& VO and Rayleigh scattering opacities}
We now include Rayleigh scattering of H$_2$, He, CO$_2$, CO, CH$_4$ and H$_2$O. For the cross-sections 
we use the values reported in \citet{dalgarnowilliams1962} (H$_2$),
\citet{chandalgarno1965} (He), \citet{sneepubachs2005} (CO$_2$, CO, CH$_4$) and \citet{harveygallagher1998} (H$_2$O).

We also calculated cross-sections for  the metal oxides TiO and VO, for which 
we used an updated line list based on \citet{plez1998}. This line list is 
available via the website\footnote{\url{http://www.pages-perso-bertrand-
plez.univ-montp2.fr/}} of Bertrand Plez for TiO. A line list obtained from 
Betrand Plez (private communication) was obtained for VO. The TiO partition 
function was taken from Uffe Gr\r{a}e J{\o}rgensen's website\footnote{\url{http://www.astro.ku.dk/~uffegj/scan/scan_tio.pdf}} for TiO and 
obtained from Betrand Plez (private communication) for VO. The VO partition 
function is based on an updated partition function by \citet{sauvaltatum1984}, 
see \citet{gustafssonedvardsson2008}.
Pressure broadening information for TiO and VO was not available. Therefore 
the broadening was approximated by use of Equation (15) in 
\citet{sharpburrows2007}.

\subsection{Scattering}
The effect of scattering on the temperature structure of the atmosphere was 
included by implementing isotropic scattering. We treat the scattering for both 
the stellar insolation and the planetary flux. For speeding up convergence we 
used accelerated lambda iteration \citep[ALI, see][]{olsonauer1986} and Ng 
acceleration \citep{ng1974}. To test our scattering implementation we 
compared the atmospheric bond albedo as a function of the incidence angle of 
the stellar light to the values predicted by Chandrasekhar's H functions 
\citep{chandrasekhar1950} and found excellent agreement.

The validity of isotropic scattering, especially for larger cloud particles, is 
questionable, because the scattering anisotropy $g$ for the cloud particles 
can be different from 0. The particles then have a strong $\sim$forward 
scattering component, such that the effective scattering opacity is smaller. The 
scattering anisotropy is defined as
\beq
g = \int_{4 \pi} (\mathbf{n}'\cdot\mathbf{n}) p(\mathbf{n}',\mathbf{n}) d{\Omega 
'} \ ,
\eeq
where $p(\mathbf{n}',\mathbf{n})$ is the scattering phase function, with 
$p(\mathbf{n}',\mathbf{n}) d{\Omega '}$ being the probability to scatter 
radiation traveling in direction $\mathbf{n}'$ into direction $\mathbf{n}$. For 
diffusive radiation it can be shown that the effective scattering opacity, the so-called reduced opacity $\kappa^{\rm scat}_{\rm red}$, can be written as 
\citep[see, e.g.][]{wang2007biomedical}
\beq
\kappa^{\rm scat}_{\rm red} = (1-g)\kappa^{\rm scat} \ ,
\eeq
where $\kappa^{\rm scat}$ is the standard scattering opacity. This result is 
independent of the actual analytic form of $p$. Therefore, in order to 
approximate the anisotropy (i.e. forward scattering) of the cloud scattering 
process, and to recover the correct anisotropic scattering in the diffusive limit, 
we take the common approach of using the reduced scattering opacity in all 
our scattering calculations. Note that for Rayleigh scattering it holds that 
$g_{\rm Rayleigh}=0$ as $p_{\rm Rayleigh} = [3/(16\pi)](1+{\rm cos}^2\theta)$. 
The symmetry $p_{\rm Rayleigh}(\theta)=p_{\rm Rayleigh}(-\theta)$ means 
that for Rayleigh scattering the fractions of forward and backward scattered 
light are equal.

\subsection{Transmission spectroscopy}
We have extended our code to also calculate transmission spectra. To obtain 
the spectra we directly calculate the transmission of stellar light through 
atmospheric annuli in transits geometry. We combine all annuli's transmissions 
which results in an effective planetary radius. We verified our implementation 
by comparing to the 1-d transmission spectra shown in figures 2 and 3 in 
\citet{fortneyshabram2010} and found very good agreement.

\section{Derivation of the \citet{ackermanmarley2001} cloud model}
\label{app:cloud_deriv}
In this section we derive the cloud density equation used in 
\citet{ackermanmarley2001}. The goal is to vertically solve for the cloud 
density of a given species. As will be shown it turns out that this equation is 
completely independent of any nucleation, condensation, coagulation, 
coalescence and shattering processes, except for the mass averaged settling 
speed in a given layer.

We define the mass fraction of cloud particles with radius $r \in [a,a+da]$ as 
$X_{\rm c}'(a)da$.
The evolution of the cloud particle mass fraction per unit radius can then be 
described by
\beq
\rho\frac{\partial X_{\rm c}'}{\partial t} = \frac{\partial }{\partial z}\left(K\rho
\frac{\partial X_{\rm c}'}{\partial z}\right)-\frac{\partial }{\partial z}\left(v_{\rm f}
\rho X_{\rm c}'\right)+S_{\rm cond+nuc}'+S_{\rm coag+dest}' \ ,
\label{equ:cloud_std}
\eeq
where $\rho$ is the total atmospheric density, $K$ the eddy diffusion constant, 
$v_f(r)$ is the particle settling speed,
$S_{\rm cond+nuc}'$ is the amount of mass added to particles in the radius 
bin due to condensation and nucleation and $S_{\rm coag+dest}'$ is the gain 
and loss of particles in the radius bin due to coagulation and collisional 
shattering. See \citet{agundezparmentier2014} (and the references therein) for 
a motivation of the functional form of the cloud particle diffusion term (first term 
on the RHS of Equation \ref{equ:cloud_std}).

In order to obtain the time derivative of the cloud particle mass density $
\rho_{\rm c} = X_{\rm c}\rho$ we set
\beq
\frac{\partial \rho_{\rm c}}{\partial t} = \rho \frac{\partial X_{\rm c}}{\partial t}= 
\rho\int\frac{\partial X_{\rm c}'}{\partial t}da \ ,
\eeq
where it was used that $\rho$ is constant in time (hydrostatic equilibrium). The 
unprimed $X$ (instead of $X'$) is the mass fraction integrated over particle 
radius. This yields 
\beq
\frac{\partial \rho_{\rm c}}{\partial t} = \frac{\partial }{\partial z}\left(K\rho
\frac{\partial X_{\rm c}}{\partial z}\right)-\frac{\partial }{\partial z}\left(\left<v_{\rm 
f}\right>\rho X_{\rm c}\right)+S_{\rm cond+nuc} \ ,
\label{equ:c_evo}
\eeq
where the coagulation/destruction term vanishes because it does not add or 
remove any mass in the condensed phase of the cloud species. $\left<v_{\rm 
f}\right>$ is the cloud particle mass averaged settling velocity.

For the gas phase of the cloud species the change in density works out to be
\beq
\frac{\partial \rho_{\rm g}}{\partial t} = \frac{\partial }{\partial z}\left(K\rho
\frac{\partial X_{\rm g}}{\partial z}\right)-S_{\rm cond+nuc} \ ,
\label{equ:g_evo}
\eeq
where settling of the gas molecule has been neglected; to first order gravity is 
balanced by the pressure gradient in the atmosphere.
To obtain the total density evolution of the cloud species within the 
atmosphere we add equations \ref{equ:c_evo} and \ref{equ:g_evo} to obtain
\beq
\frac{\partial \rho_{\rm g+c}}{\partial t} = \frac{\partial }{\partial z}\left(K\rho
\frac{\partial X_{\rm g+c}}{\partial z}-\left<v_{\rm f}\right>\rho X_{\rm c}\right)
\eeq
For a steady state solution and a net zero mass flux one thus finds that
\beq
K\frac{\partial X_{\rm g+c}}{\partial z}-\left<v_{\rm f}\right> X_{\rm c} = 0
\eeq
which is equal to the equation solved in \citet{ackermanmarley2001} except 
for the fact that they express the settling speed in units of the convective 
mixing speed.

If one wants to solve for the condensate mass fraction, assuming that the gas 
mass fraction is known then one finds from the previous equation that
\beq
K\frac{\partial X_{\rm c}}{\partial z} - \left<v_{\rm f}\right>X_{\rm c} = - K
\frac{\partial X_{\rm g}}{\partial z} \ .
\eeq
In the case of effective heterogeneous nucleation, i.e. if the nucleation 
timescale is shorter than the convective mixing or settling timescale, it can be 
assumed that $X_{\rm g}$ is a known quantity:
\beq
X_{\rm g} = X_{\rm s} \ ,
\eeq
where $X_{\rm s}$ is the saturation gas mass fraction and one finds
\beq
K\frac{\partial X_{\rm c}}{\partial z} - \left<v_{\rm f}\right>X_{\rm c} = - K
\frac{\partial X_{\rm s}}{\partial z} \ .
\eeq
In this case $X_{\rm s}$ can be obtained from an equilibrium chemistry 
module. For species such as MgSiO$_3$ no gaseous phase exists because 
the molecule forms through chemical nucleation, therefore $\partial X_{\rm s}/
\partial z$ cannot be calculated using equilibrium chemistry.
However, $-\partial X^{\rm equ}_{\rm c}/\partial z$, i.e. the negative gradient of 
the equilibrium chemistry condensate mass fraction, is a measure for the gas 
mass gradient between two layers which arises due to the chemical nucleation 
of various gaseous species in order to form the condensate (for MgSiO$3$ it's 
Mg(g), SiO(g) and O$_2$(g)). This yields
\beq
K\frac{\partial X_{\rm c}}{\partial z} - \left<v_{\rm f}\right>X_{\rm c} = - K
\frac{\partial X_{\rm c}^{\rm equ}}{\partial z} \ .
\label{equ:cloud_struct_final}
\eeq
For species which can exist in the gas phase, and for which the condensation occurs
on a small enough spatial range such that the total chemical 
abundances (gas+solid) are roughly constant, one can assume that $\partial X_{\rm t}
^{\rm equ} / \partial z = 0$, with $X_{\rm t}^{\rm equ} = X_{\rm c}^{\rm equ} + 
X_{\rm s}$. This leads to $\partial X_{\rm s} / \partial z=-\partial X_{\rm c}^{\rm 
equ} / \partial z$, such that we will use Equation (\ref{equ:cloud_struct_final}) 
in all cases.

A test carried out for a self-luminous planet with $T_{\rm eff} = 500 K$, set up 
identically as in \citet{ackermanmarley2001}, but using our form of the cloud 
equation as written in Equ. \ref{equ:cloud_struct_final} yielded very good 
agreement.

The particle sizes for the clouds are calculated as described in \citet{ackermanmarley2001}.

\bibliographystyle{aa}
\bibliography{mybib}{}

\end{document}